\documentclass[a4wide]{article}
\usepackage{latexsym}
\usepackage{amsmath}
\usepackage{amssymb}
\usepackage{listings}
\usepackage{graphicx}
\usepackage{mydefs}
\usepackage{xspace}
\usepackage{url}
\usepackage{geometry}
\usepackage{mathptmx}
\newcounter{lst}[section]

\lstdefinestyle{easslisting}{basicstyle=\footnotesize\sffamily, mathescape=true, frame=tb,  numbers=right, numberstyle=\footnotesize, stepnumber=1, numbersep=-5pt, captionpos=b}

\lstdefinestyle{eass}{basicstyle=\sffamily, mathescape=true}

\lstnewenvironment{listing}[3]{
  \noindent        
  \refstepcounter{lst}         
  \label{code:#1}  
\begin{tabular}{p{.97\columnwidth}} \\ \hline  {\normalsize \textbf{Code  fragment \arabic{section}.\arabic{lst}} #2} \\  \end{tabular} 
\lstset{language=#3,          
  basicstyle=\footnotesize\sffamily,
  xleftmargin=10pt,    
  mathescape=true,    
  frame=tb,    
  numbers=right,    
  numberstyle=\footnotesize,     
  stepnumber=1,     
  numbersep=-5pt}}{}

\lstdefinelanguage{Gwendolen}{%
    morekeywords={Plans,Initial,Beliefs,Goals,name,fof-parse,Rules,Belief},
    morecomment=[l]{//},
 literate= {<-}{{$\leftarrow$}{$\:$}}2
           {.B}{{${\cal B}$}}2
           {.G}{{${\cal G}$}}2
           {lnot}{{$\sim$}}2
           {assert_shared}{{$+_{\Sigma}$}}2
           {remove_shared}{{$-_{\Sigma}$}}2
           {True}{{$\top$}}2
           {(perform)}{}0
}

\newcommand{\mahca}{MAHCA}
\newcommand{\keymaera}{KeYmaera}
\newcommand{\mcapl}{MCAPL\xspace}
\newcommand{\mck}{MCK}
\newcommand{\ajpf}{AJPF\xspace}
\newcommand{\jpf}{JPF\xspace}
\newcommand{\java}{\texttt{Java}}
\newcommand{\ail}{AIL\xspace}
\newcommand{\gwendolen}{\textsc{Gwendolen}}

\newcommand{\matlab}{\textsc{MatLab}}
\newcommand{\simulink}{\textsc{Simulink}}

\newcommand{\jason}{\textit{Jason}}
\newcommand{\agentspeak}{\textit{AgentSpeak}}
\newcommand{\jadex}{\textit{Jadex}}
\newcommand{\goal}{\textsc{GOAL}}

\newcommand{\psl}{PSL\xspace}
\newcommand{\tripleapl}{\textsc{3APL}}
\newcommand{\mcmas}{MCMAS}
\newcommand{\hytech}{\textsc{HyTech}}
\newcommand{\phaver}{\textsc{PHAVer}}
\newcommand{\lustre}{\textsc{Lustre}}
\newcommand{\brahms}{\textit{Brahms}}
\newcommand{\spin}{SPIN}

\newcommand{\prism}{PRISM}
\newcommand{\AILagent}{\ensuremath{ag}}

\newcommand{\lbelief}[2]{\ensuremath{\mathrm{\mathbf{B}}_{#1} \, #2}}

\newcommand{\lgoal}[2]{\ensuremath{\mathrm{\mathbf{G}}_{#1} #2}}

\newcommand{\lactions}[2]{\ensuremath{\mathrm{\mathbf{A}}_{#1} #2}}
\newcommand{\lactionsfunc}{\ensuremath{\mathrm{\mathbf{A}}}}

\newcommand{\lintention}[2]{\ensuremath{\mathrm{\mathbf{I}}_{#1} #2}}
\newcommand{\lintentionfunc}{\ensuremath{\mathrm{\mathbf{I}}}}

\newcommand{\lpercept}[1]{\ensuremath{\mathrm{\mathbf{P}}(#1)}}
\newcommand{\lperceptfunc}{\ensuremath{\mathrm{\mathbf{P}}}}
\newcommand{\MAS}{\ensuremath{\mathit{MAS}\xspace}}

\def\sometime{\lozenge}
\def\always{\square}
\newcommand{\until}{\ensuremath{\mathop{\mathrm{\mathcal{U}}}}}
\newcommand{\release}{\ensuremath{\mathop{\mathrm{\mathcal{R}}}}}
\newcommand{\tlnext}{\!\raisebox{-.2ex}{ 
                        \mbox{\unitlength=0.9ex
                        \begin{picture}(2,2)
                        \linethickness{0.06ex}
                        \put(1,1){\circle{2}} 
                        \end{picture}}}       
                        \,}

\let\imp=\Rightarrow

\let\imp=\Rightarrow
\def\llor{\,\lor\,}

\begin{document}
\lstset{language=gwendolen,style=eass}
\title{Practical Verification of Decision-Making in 
       Agent-Based Autonomous Systems}

\author{Louise A.~Dennis \\
L.A.Dennis@liverpool.ac.uk,  Department of Computer Science, University of Liverpool, UK 
\and Michael Fisher \\
Department of Computer Science, University of Liverpool, UK
\and Nicholas K.~Lincoln
\and Alexei Lisitsa \\
Department of Computer Science, University of Liverpool, UK
\and Sandor M.~Veres \\
Department of Automatic Control and Systems Engineering, University of Sheffield, UK}


\maketitle

\begin{abstract}
  We present a verification methodology for analysing the
  decision-making component in agent-based hybrid systems.
  Traditionally hybrid automata have been used to both implement and
  verify such systems, but hybrid automata based modelling,
  programming and verification techniques scale poorly as the
  complexity of discrete decision-making increases making them
  unattractive in situations where complex logical reasoning is
  required.  In the programming of complex systems it has, therefore,
  become common to separate out logical decision-making into a
  separate, discrete, component. However, verification techniques have
  failed to keep pace with this development.  We are exploring
  agent-based logical components and have developed a model checking
  technique for such components which can then be composed with a
  separate analysis of the continuous part of the hybrid system.
  Among other things this allows \emph{program} model checkers to be
  used to verify the actual implementation of the decision-making in
  hybrid autonomous systems.
\end{abstract}

\setcounter{tocdepth}{1}
\tableofcontents
\newpage

\section{Introduction}
\label{sec:introduction}
Autonomous systems are moving beyond industrial and military contexts,
and are now being deployed in the home, in health-care scenarios, and
in automated vehicles. In many cases there may be a human operator who
directs the autonomous system, but increasingly the system must work
on its own for long periods without such interventions. These systems
must essentially \emph{decide for themselves} what to do and when to
do it. This might seem fraught with danger, but there is a clear need
for such systems, particularly:
\begin{itemize}
\item when deployed in \emph{remote} or \emph{dangerous} environments
  where direct and local human control is infeasible; or
\item when the \emph{complexity} or \emph{speed} of the environmental
  interactions is too high for a human to handle.
\end{itemize}
Examples of the former include deep sea exploration, space probes, and
contaminated area cleanup; examples of the latter include automated
stock trading systems, robot swarms, and unmanned air vehicle
collision avoidance. In addition to the above reasons, autonomous
systems are becoming popular as they can sometimes be much
\emph{cheaper} to develop and deploy than manned systems.

\subsection{Hybrid Autonomous Systems}
Autonomous systems are increasingly constructed from distinct
components, some handling continuous interaction, some dealing with
discrete analysis. Traditionally, such hybrid autonomous systems have
been engineered using the concept of a hybrid automaton.  However, as
these systems have become more complex, combining discrete
decision-making and continuous control within a hybrid automaton has
faced challenges.  It is difficult to separate the two concerns
(decision-making and continuous control) when designing and
implementing a system in this fashion -- this has an impact on design
and code, understandability and reuse.  Furthermore many autonomous
systems operate in environments where the users wish to access some
high level account for \emph{why} a decision was taken by the
system~\cite{FDW12:magazine}, such an account can be difficult to
extract from a hybrid automaton.

As a result, a distinct strand of research has focused on explicitly
separating decision-making components from the underlying control
system, often using the popular agent
paradigm~\cite{Wooldridge02:book}.  A drawback of this approach,
however, is that it is generally non-trivial to transform such a
system back into a hybrid automaton based model and so well-developed
techniques for verifying hybrid automata by model
checking~\cite{HenzingerHW97,frehse05:_phaver} become difficult to
apply to these new systems.  Moreover, verification of hybrid
automaton based systems tend to scale badly as the reasoning processes
become more complex.  Since autonomous systems are frequently safety
or mission critical this verification gap is a significant concern.

\subsection{A Methodology for Verifying Autonomous Choices}
\label{sec:choices}
In~\cite{FDW12:magazine} we highlighted a methodology for the
verification of decision-making components in hybrid autonomous
systems, where such a decision-making component is implemented as a
rational agent. In this paper, we give a more detailed and technical
explanation of the methodology and apply the approach to practically
verify a variety of complex autonomous systems (see
sections~\ref{sec:scen2}, \ref{sec:eass:arch}, and
\ref{sec:adaptive}). We argue that the most crucial aspect of
verifying complex decision-making algorithms for autonomous systems,
for example concerning safety, is to identify that the controlling
agent \emph{never} deliberately makes a choice it \emph{believes} to
be unsafe\footnote{Further discussion of this aspect is provided
  in~\cite{FDW12:magazine}.}.  In particular this is important in
answering questions about whether a decision-making agent will make
the same decisions as a human operator given the same information from
its sensors.

Thus, rather than verifying agent behaviour within a detailed model of
the system's environment, we will now verify the choices the agent
makes given the beliefs it has. This approach is clearly simpler than
monolithic approaches as we can state properties that only concern the
agent's internal decisions and beliefs, and so verification can be
carried out without modelling the ``real world''.  At a logical level,
the verification of \emph{safety} properties changes from the global
checking of\footnote{`$\always$' and `$\sometime$' are temporal logic
  operators meaning ``at all future moments'' and ``in some future
  moment'', respectively, while `$\lbelief{agent}$' is a logical
  operator describing the beliefs the agent has.}
$$\always\lnot\mathsf{bad}$$
i.e., nothing bad can ever happen, to locally
checking\footnote{Alternatively:
  $\lbelief{agent}{\always\lnot\mathsf{bad}$} if the agent can hold
  temporal beliefs.}
$$\always\lbelief{agent}{\lnot\mathsf{bad}}$$
i.e., the agent never believes that something bad happens. Similarly,
\emph{liveness} properties change from checking $\sometime
\mathsf{good}$ globally to checking $\sometime
\lbelief{agent}{\mathsf{good}}$ locally. Thus, we verify the (finite)
\emph{choices} the agent has, rather than all the (typically infinite)
``real world'' effects of those choices.

Specifically, we propose \emph{model checking} (and the \emph{model
  checking of programs}, if possible) as an appropriate tool for
demonstrating that the core rational agent always endeavours to act in
line with our requirements and never \emph{deliberately} chooses
options that lead internally to bad states (e.g., ones where the agent
believes something is unsafe).  Since we are verifying only the core
agent part we can use non-hybrid model checking approaches which
allows the verification to be more easily compositional and gives us a
wider range of abstraction mechanisms to explore.  Thus, we do not
verify all the ``real world'' outcomes of the agent's choices (but
assume that analysis of the underlying control system has provided us
with theorems about the outcomes of actions etc.), but do verify that
it always tries to achieve its goals/targets to the best of its
knowledge/beliefs/ability. Thus, the agent \emph{believes} it will
achieve good situations and \emph{believes} it will avoid bad
situations. Consequently, any guarantees here are about the autonomous
system's decisions, not about its external effects.

\subsection{Autonomy in the Real World}
Autonomous systems, such as we are discussing, are constructed in a
component-based fashion with an agent-based decision maker and a
control system.  This means they have a natural decomposition at
design time.  In many cases the agent-based decision maker is viewed
as the replacement for a human pilot or operator who would, otherwise,
interact with the control system.  

How might we \emph{guarantee} behaviour within such an autonomous
system embedded in the real world? Following a decompositional
approach we need \emph{not} (and, indeed we argue that we can not, for
the reasons outlined in Section~\ref{sec:real_world}) ensure that an
autonomous system will \emph{certainly} lead to a change in the real
world.  When using the model-checking approach to verification,
efforts to model the entire system and its interaction with the real
world with any degree of accuracy necessarily involve complex
abstractions together with a number of assumptions.  These
abstractions and assumptions are embedded deep within an executable
model and may not be explicit to end users, or even to the
modellers. Therefore if we provide a guarantee, for example, that the
autonomous system can definitely achieve or avoid something, there
will be a number of pre-conditions (that the real world will behave in
some particular way) to that guarantee that may be hard to extract.
One of the aims of our approach is that the assumptions embedded in
the modelling of the real world should be as explicit as possible to
the end users of a verification attempt.

Obviously, some parts of an agent's reasoning are triggered by the
arrival of information from the real world and we must deal with this
appropriately. So, we first analyse the agent's program to assess
what these incoming \emph{perceptions} can be, and then explore, via
the model checker, all possible combinations of these.  This allows us
to be agnostic about how the real world might actually behave and
simply verify how the agent behaves \emph{no matter what} information
it receives. Furthermore, this allows us to use hypotheses that
explicitly describe how patterns of perceptions might occur. Taking
such an approach clearly gives rise to a large state space because we
explore all possible combinations of inputs to a particular agent.
However it also allows us to investigate a multi-agent system in a
compositional way. Using standard \emph{assume-guarantee} (or
\emph{rely-guarantee})
approaches~\cite{MisraC81,Jones83,Jones86a,MannaPnueli92:book,Lamport03:TLA},
we need only check the internal operation of a single agent at a time
and can then combine the results from the model checking using
deductive methods to prove theorems about the system as a whole.
Abstracting away from the continuous parts of the system allows us to
use model checking in a compositional fashion.

It should be noted that, in many ways, our approach is the complement
of the typical approach employed in the verification of hybrid
automata and hybrid programs.  We are primarily concerned with the
correctness of the discrete algorithms and are happy to abstract away
from the underlying continuous system, while the other approaches are
more concerned with the verification of the continuous control and are
happy to abstract away from the discrete decision-making algorithms.

\subsection{Overview}
In summary, we had two main aims in developing our verification
methodology: to maintain the natural decomposition of the system
design in the verification, allowing us to verify the symbolic and
non-symbolic aspects separately; and to make any assumptions about the
continuous/real world environment as explicit as possible. It is
preferable to treat the verification and analysis in a compositional
fashion, if at all possible, for a variety of reasons. It simplifies
the verification task, encourages reuse, and allows appropriate domain
specific analysis tools to be used.  Moreover, in many of the
situations we are interested in, the control system \emph{is} trusted
and the primary concern is the extent to which the decision-making
component conforms to the expectations about the behaviour of a human
operator.  Therefore we have an additional motivation to verify the
symbolic reasoning in detail and in isolation.
\medskip

\noindent Thus, in summary, our overall approach~\cite{FDW12:magazine}
involves:
\begin{enumerate}
\item modelling/implementing the agent behaviour and describing the interface
  (input/output) to the agent;
\item model checking the decision-making agent within an coarse
  over-approximation of the environment derived from the analysis of
  agent inputs (this will establish some property, $\varphi$);
\item if available, utilizing environmental hypotheses/assumptions, in
  the form of logical statements, to derive further properties of the
  system;
\item if the agent is refined, then modify (1) while if environmental
  properties are clarified, modify (3); and
\item deducing properties of multi-agent systems by model checking the
behaviour of individual agents in a component-wise fashion and then
combining the results deductively to infer properties of the whole
system.
\end{enumerate}
We will later examine three very different
scenarios in order to exemplify the key features of the methodology.

\section{Background and Related Work}
The verification of hybrid agent systems, where decision-making takes
place as part of a separate agent-based component, draws upon
background and research from a range of areas, namely hybrid control
systems, agent-based programming and verification approaches to hybrid
systems and rational agents.

\subsection{Hybrid Control Systems}

A fundamental component of low-level control systems technology is the
\emph{feedback controller}. This measures, or estimates, the current
state of a system through a dynamic model and produces subsequent
feedback/feed-forward control signals. In many cases, difference
equations can be used to elegantly manage the process. These equations
of complex dynamics not only make changes to the input values of
sub-systems and monitor the outcomes on various sensors, but also
allow deeper understanding of system behaviour via analytical
mathematical techniques.

As we move to autonomous systems, such controllers are increasingly
required to work in situations where there are areas of discontinuity,
where a distinct change in behaviour is required and where control
needs to switch to the use of an alternative model and alternative
control equations. This kind of hybrid control
system often requires some
decision-making system to be integrated with the feedback
controller~\cite{BBM98,DesHybridSysts99,HyDySy09}. It may also be necessary for a system to take actions such
as detecting that a fuel line has ruptured and switching valves to
bring an alternative online: this falls outside the scope of monitoring
and adjusting input and output values, and involves detecting that
thresholds have been exceeded or making large changes to the system.

Although it is possible to encode all these choices in a larger, often
hierarchical, control system, based on hybrid automata this creates
several problems from a software engineering perspective.  Firstly, it
does not allow a natural {\em separation of concerns} between the
algorithms and procedures being designed or implemented for making
decisions, and those which are being designed and implemented for
continuous control, this reduces the {\em understandability} of the
resulting code and therefore increases the risk of errors and bugs
being present in the code.  It can often also make the underlying
code, particularly that of decision-making algorithms, harder to {\em
  reuse} since it becomes more difficult to identify the generic
aspects of any decision algorithm from those that are specific to the
continuous control.  There is also
evidence~\cite{damm07:_exact_state_set_repres_verif,DFLLV10:iSairas}
that hybrid automata based implementations scale poorly with the
complexity of decision-making when compared to agent-based control.
This again impacts upon understandability.

\begin{figure}
\begin{center}
\includegraphics[width=0.889479\textwidth]{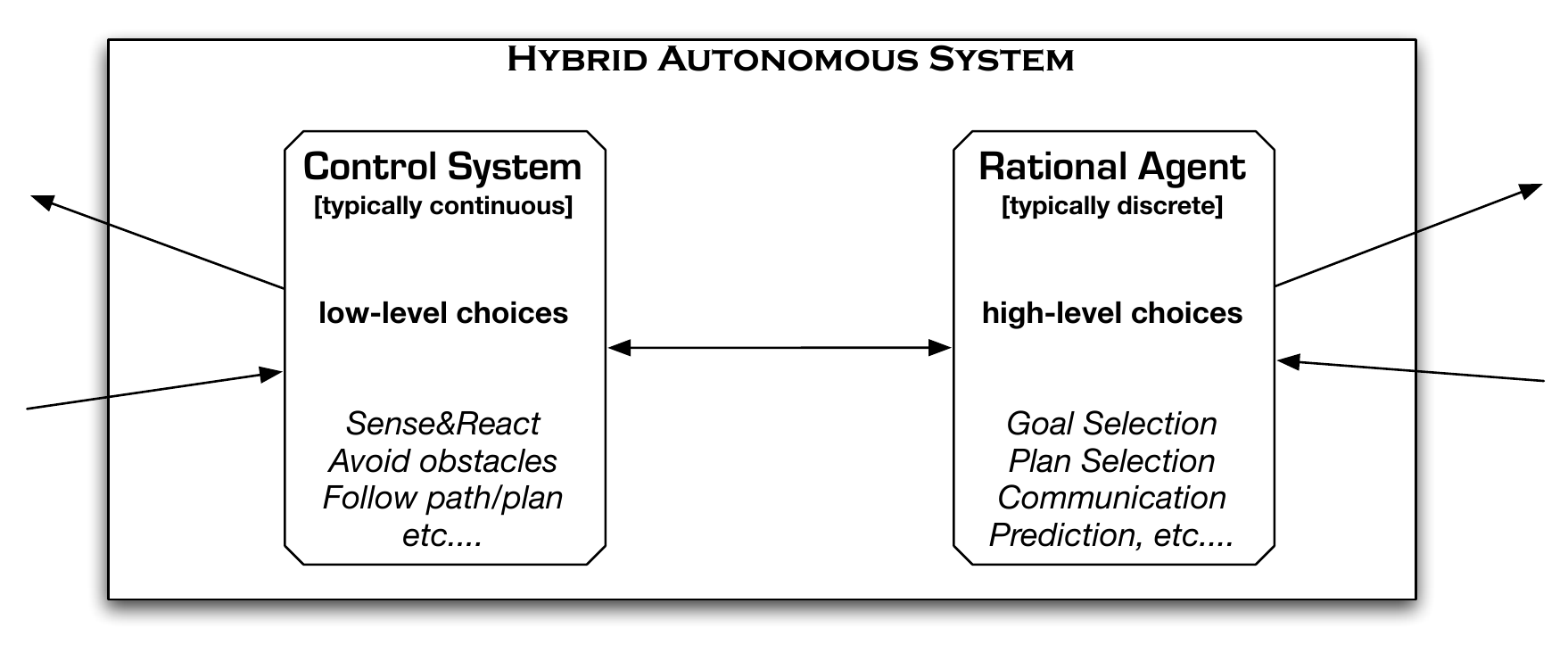}
\end{center}
\caption{The Typical Structure of a Hybrid Autonomous System.  Low
  level choices such as (e.g., tight sense-react loops) are governed
  by the control system while high-level choices (e.g., managing long
term goals) are governed by a rational agent.}
\label{fig:hybrid2}
\end{figure}

Systems which explicitly separate logical decision-making from continuous control are often referred to as
\emph{hybrid control systems}, in that they integrate discrete,
logical decision processes with physical system dynamics. In the case
of autonomous systems, the control system usually follows fairly well
defined functions, while the discrete decision-making process must
make appropriate choices, often without human intervention; see
Fig.~\ref{fig:hybrid2}.  Increasingly, this discrete decision-making
process will be a \emph{rational agent}, able to make justifiable
decisions, to reason about those decisions, and dynamically modify its
strategy~\cite{WooldridgeRao99:book}.

\subsection{Agents and Rational Agents}
At its most general, an \emph{agent} is an abstract concept that
represents an \emph{autonomous} computational entity that makes its
own decisions~\cite{Wooldridge02:book}.  Since its introduction in the
1980s, the agent abstraction has become very influential in practical
situations involving complex flexible, autonomous, and distributed
components~\cite{bond:88a,bratman:88a,davis:83a,cohen:90a,durfee:89d,shoham:93a}.

A general agent is simply the encapsulation of some distributed computational component within a larger system.  However, in many settings,
something more is needed. Rather than just having a
system which makes its own decisions in an opaque way, it is
increasingly important for the agent to have explicit \emph{reasons}
(that it could explain, if necessary) for making one choice over
another.  

In the setting of autonomous systems, explicit reasoning assists
acceptance.  When queried about its choices an autonomous system
should be able to explain them, thus allowing users to convince
themselves over time that the system is reasoning competently.  More
importantly, where an autonomous system is likely to be the subject of
certification, it may need to fulfill obligations to do with
reporting, particularly in instances where an accident or ``near
miss'' has occurred.  It is desirable for such reporting to be
presented at a high level which explains its choices in terms of what
information it had, and what it was trying to achieve.

\emph{Rational} agents~\cite{bratman:87a,rao:92a,WooldridgeRao99:book} enable the representation of this kind of reasoning.
Such an agent has explicit reasons for
making the choices it does.  We often describe a rational agent's \emph{beliefs} and \emph{goals}, 
which in turn determine the agent's 
\emph{intentions}. Such agents make decisions about what action to
perform, given their current beliefs, goals and intentions.

\subsubsection{BDI Agents and Programming Languages}
The predominant view of rational agency is that encapsulated within
the BDI model~\cite{rao:91c,rao:92a,rao:95b}. Here, `BDI' stands for
\emph{Beliefs}, \emph{Desires}, and \emph{Intentions}. Beliefs
represent the agent's (possibly incomplete, possibly incorrect)
information about itself, other agents, and its environment, desires
represent the agent's long-term goals while intentions represent the
goals that the agent is actively pursuing.

There are \emph{many} different agent programming languages and agent
platforms based, at least in part, on the BDI approach. Particular
languages developed for programming \emph{rational} agents in a
BDI-like way include \agentspeak~\cite{rao96:agentspeak},
\jason~\cite{MAPlpaX:Bordini,Book:JASON},
\tripleapl~\cite{3APL98,MAPlpaX:Dastani},
\jadex~\cite{PokahrBL05},
\brahms~\cite{sierhuis01:_model_simul_work_prati},
\goal~\cite{GOAL01,boer07}, and \gwendolen~\cite{dennis08:_gwend}.
Agents programmed in these languages commonly contain a set of
\emph{beliefs}, a set of \emph{goals}, and a set of \emph{plans}.
Plans determine how an agent acts based on its beliefs and goals and
form the basis for \emph{practical reasoning} (i.e., reasoning about
actions) in such agents.  As a result of executing a plan, the beliefs
and goals of an agent may change as the agent performs actions in its
environment.
\bigskip

\subsubsection{Agent Controlled Hybrid Systems}

Agent based control of decision-making in hybrid systems has been an
area of research since at least the 1990s.  Kohn and Nerode's \mahca\
system~\cite{kohn92:_multip_agent_auton_hybrid_contr_system} uses
multiple knowledge-based agents as planners.  While these agents are
not based on the BDI paradigm, which was only in its infancy when
\mahca\ was originally developed, the approach is motivated by a
desire to represent logical decision-making in a high-level
declarative fashion.

More recently agent based approaches have been explored in the control
of spacecraft~\cite{RemoteAgent98}, Unmanned
Aircraft~\cite{Karim:2005:EDI:1082473.1082799,webster11formal,patchett10:_devel_advan_auton_integ_mission},
and robotics~\cite{wei13:_agent_based_cognit_robot_archit}.  Many of
these approaches are explicitly BDI based and are motivated by the
desire to separate the symbolic and sub-symbolic reasoning and model
the mission designer's \emph{intent}.

\subsection{Formal Verification and Model Checking}

Formal verification is essentially the process of assessing whether a
specification given in formal logic is satisfied on a particular
formal description of the system in question. For a specific logical
property, $\varphi$, there are many different approaches to
this~\cite{Fetzer88:veryidea,dMLP79,BoyerMoore81}, ranging from 
deductive verification against a logical description of the system
$\psi_S$ (i.e., $\vdash \psi_S\imp\varphi$) to the algorithmic
verification of the property against a model of the system, $M$
(i.e., $M\models\varphi$). The latter has been extremely successful in
Computer Science and Artificial Intelligence, primarily through the
\emph{model checking} approach~\cite{Clarke00:MC}. This takes an executable model
of the system in question, defining all the model's possible
executions, and then checks a logical property against this model
(and, hence, against all possible executions).

Whereas model checking involves assessing a logical specification
against all executions of a \emph{model} of the system, an alternative
approach is to check a logical property directly against all
\emph{actual} executions of the system. This is termed the \emph{model
  checking of programs}~\cite{VisserHBPL03} and crucially depends on
being able to determine all executions of the actual program. In the
case of \java{}, this is feasible since a modified \emph{virtual
  machine} can be used to manipulate the program executions. The Java
Pathfinder (\jpf) system~\cite{VisserHBPL03} carries out formal
verification of \java{} programs in this way by exhaustively analysing
all the possible execution paths. This avoids the need for an extra
level of abstraction and ensures that verification truly occurs on the
real system. However, although exhaustive, such techniques are
typically \emph{much} slower than standard model checking.

\subsection{Verifying Hybrid Systems}
As  mentioned above, there are two major approaches to constructing hybrid systems.  We
have followed the methodology of providing two components, a decision
making system that interacts with an underlying control system.  The
other major approach is that of hybrid automata.  

An automaton consists of a set of states with transitions between
them.  Traditionally automata states are static but, in hybrid
automata, the states are governed by differential inclusions which
control the behaviour of the overall system while the automata is in
that state.  Each state therefore contains (state) \emph{variables}
whose value may evolve and change in a continuous fashion while the
automata is in that state.  Key aspects of a hybrid automaton state
are the \emph{flow conditions}; equations which express the evolution
of continuous variables while in that state, and \emph{invariant
  conditions}; a predicate over the state variables which must be
satisfied in that state.  If the invariant conditions are violated
then the state is forced to transition to another.  When a transition
occurs, \emph{jump conditions} determine the initial values of the
state variables in the next state.

The model checking of hybrid automata involves exploring
\emph{trajectories} within each automata state.  Given (a set of)
starting values for the variables in the state the system attempts to
determine a bounded sub-region of $\mathbb{R}^n$ which contains all
possible values of the variables that can be reached within that state
before it transitions to another.  The possible values of the
variables at the point of transition, together with the jump
conditions, will form the starting values for the variables in the
next state, and so on until all possible states of the system have
been
explored. (See~\cite{Henzinger96thetheory,AlurDiscreteAbstractions00}
and \cite{HenzingerHW97} for detailed descriptions of the model
checking of hybrid automata based systems.)

From our perspective: the verification of agent-based hybrid control
systems, with a particular interest in the verification of
implementations of decision-making algorithms, the hybrid automata
approach has several drawbacks, as follows.
\begin{enumerate}
\item As noted earlier, it is difficult to decompose a hybrid
  automaton with a complex decision-making process into sub-systems
  (though approaches using concurrent automata are possible).
  However, even after such a decomposition, the model checking problem
  can not be easily decomposed to consider only some sub-set of the
  states of the system (e.g., just those involved with control
  algorithms).  The calculation of a bounded region that limits the
  trajectories within a state is dependent upon the possible values of
  the variables on entering the state and these, in turn, depend upon
  the values of those variables in the previous state and the jump
  conditions.  Therefore the region of possible trajectories may vary
  each time a state is visited.  As a result, the algorithm needs to
  calculate all the potential sets of starting values for each state.
  It is not, in general, possible to determine in advance what the
  entry constraints will be for some sub-set of the automaton states
  without exploring the whole automaton.

  As a result the whole system must often be treated as a unit. This
  not only makes the model checking problem harder but also limits the
  reuse of results when changes are made to the system.

\item The classes of hybrid automata that have decidable model
  checking problems are limited.  While there are several possible
  decidable fragments, linear hybrid automata are most frequently used
  (for instance the \hytech\ model checker works on linear hybrid
  automata~\cite{HenzingerHW97}).  The main restrictions on linear
  hybrid automata are that the formul\ae\ describing the evolution of
  dynamic variables, their constraints, and the evolution to new
  discrete states must be finite conjunctions of linear inequalities,
  and the flow conditions may refer only to derivatives of state
  variables and not to the variables themselves.  From the perspective
  of complex decision-making the restriction to conjunctions of
  inequalities, rather than more expressive formul\ae\, forces logical
  reasoning to be modelled using sequences of discrete transitions
  where intuitively only one transition should take place.

\item This in turn means that tools such as
  \hytech~\cite{HenzingerHW97}, and \phaver~\cite{frehse05:_phaver}
  implemented for the model checking of linear hybrid automata also do
  not scale well in the presence of large discrete state
  spaces~\cite{damm07:_exact_state_set_repres_verif}.  Not only do
  complex logical computations have to be modelled as a set of states
  with the same continuous dynamics, but also the process of
  determining trajectories through individual discrete states tends to
  create multiple states in the model checking algorithm for each
  state in the hybrid automata (i.e., a state for each possible region
  of trajectories through the state).

\item As noted above, a hybrid automaton models the behaviour of an
  entire system; both the computational and the real world parts. As a
  result, in general, hybrid automata do not provide a suitable
  implementation language for a programmer wishing to create the
  computational parts of some larger system.  Consequently, hybrid
  automata model checking operates much like standard model checking,
  i.e., it verifies a \emph{model} of the implemented system, rather
  than the system itself.

  Particularly in certification contexts there is a need to verify the
  \emph{actual} implemented system.  In these cases it is necessary to
  somehow compile the implemented system, together with a model of its
  environment, into an explicit hybrid automaton before analysis.
\end{enumerate}

\subsubsection{Approaches to Compositionality and Expressivity}

\phaver~\cite{frehse05:_phaver} uses assume-guarantee style reasoning
based on I/O Automata to allow the model checking of linear hybrid
automata to be approached in a compositional
fashion~\cite{frehse04:_assum_i_o}.  The approach takes a hybrid
automaton that has been constructed from the parallel composition of
smaller sub-systems (e.g., a physical system with its continuous
dynamics, and a software controller). This work describes how, during
model checking, these sub-systems can be replaced by abstractions
which are \emph{comparable}.  So a physical system can be model
checked, composed with an abstraction of a software controller and
vice versa.  This work provides results and algorithms for
establishing the correctness of the original system via verification
of these new systems and experimental results to show the improvements
in efficiency gained by verifying multiple simple systems in place of
a single monolithic complex system.  The more abstract systems
constitute an over-approximation of the original sub-system and
therefore it is possible for the verification process to return false
negatives. The underlying verification continues to use hybrid
automata based model checking and so remains constrained by many of
the considerations above (e.g., \phaver\ is restricted to linear
hybrid automata).  However the basic approach, abstracting away one
sub-system, is substantially similar to the one we adopt here.

Platzer has developed a theorem
proving approach for verifying hybrid automata which is implemented in the \keymaera{} system~\cite{Platzer10}.  This benefits from many of the advantages of theorem proving, in that it
can produce stronger, more general, results than model checking but
also suffers from its disadvantages.  Producing such verifications is
generally time-consuming and requires highly skilled users although \keymaera\ is linked to a number of automated systems for solving sub-problems.

\keymaera\ verifies hybrid systems which are represented as
\emph{hybrid programs} written in a dedicated while-style programming
language.  There is a direct embedding of hybrid automata into hybrid
programs.  However \emph{hybrid programs} are more expressive than,
for instance, linear hybrid automata, and it is possible to represent
complex logical formul\ae\ as invariants on states. Unfortunately, the
language remains at a low level, meaning that the implementation of
complex decision-making is quite cumbersome.

\subsubsection{Solutions to the Modeling Problem}

While it is theoretically possible to represent all hybrid systems as
hybrid automata or hybrid programs, it is time consuming to do so.  In
particular, as noted above, it is difficult to represent complex
computations occurring at the discrete decision-making level in a
compact fashion.  As a result, where complex decision-making takes
place a hybrid automaton or hybrid program approach frequently
abstracts away from the process and, instead, just represents all
possible outcomes of the decision-making as non-deterministic choices.
While this simplifies the hybrid automaton/program representation and
makes its creation feasible it can, as a result, introduce ``false
negatives'' into the verification process -- i.e., states exist in the
hybrid automaton/program that would not exist in reality since the
decision-making process would have vetoed them.

There have been several attempts to produce systems that will compile
higher level programming languages into hybrid automata for
verification -- thus allowing the full complexity of decision-making
to be represented.  These systems can also, potentially, employ
abstraction techniques during the compilation process in order to
reduce the search space.  Such approaches have focused on synchronous
languages such as
\lustre~\cite{briand09:_combin_contr_data_abstr_verif_hybrid_system},
and Quartz~\cite{bauer12:_new_model_languag_cyber_system}.

The agent paradigm, in particular, has been developed to address
distributed asynchronous situations which makes it attractive in
situations where a synchronous model of time is unhelpful.  However,
as yet, no system has been created to compile agent-based control into
hybrid automata.

\subsection{Model Checking Agent Programs}
\label{sec:psl}
There are a number of approaches to model checking and verifying
multi-agent systems~\cite{Alechina:11a,lomuscio09:_mcmas,gammie04:_mck}.  Most of these are not 
specifically geared towards BDI-style rational agents but provide more
general tools for the analysis of agents. In general these have not been applied to the
problem of hybrid systems although \mcmas\ has been used in the
verification of hybrid automata~\cite{ezekiel11:_verif_auton_under_vehic}.

There are three systems that focus specifically on the model checking
of BDI programs.  \cite{JongmansHR10} describe a program model
checker tailored specifically towards the \goal\ programming
language. Both~\cite{hunter13:_syner_exten_framew_multi_agent_system_verif}
and~\cite{DBLP:conf/jelia/StockerDDF12} have investigated the model
checking of \brahms~\cite{sierhuis01:_model_simul_work_prati}
programs using \spin~\cite{holzmann04spin}.  Both systems
re-implement \brahms\ in \java\ and then either export directly into
\spin~\cite{DBLP:conf/jelia/StockerDDF12} or use Java Pathfinder
(\jpf) to generate a model which is then exported to
\spin~\cite{hunter13:_syner_exten_framew_multi_agent_system_verif}.

The \mcapl\ framework~\cite{MCAPL_journal} (described in more detail
below) provides access to model checking facilities to programs
written in a wide range of BDI-style agent programming languages so
long as those languages have a \java-based program interpreter.  We
chose to use the \mcapl\ framework in this work but the general
methodology does not depend upon that choice, or even upon the choice
of a program model checker for a BDI language.

\subsubsection{The \mcapl\ Framework}
In the examples discussed later in this paper we use the
\mcapl\ framework which includes a model checker built on top of \jpf.
We do not consider the framework to be integral to our approach to the
verification of agent-based decision-making in autonomous systems,
though some general background on the system will be useful when
reading the later examples. The framework is described in detail
in~\cite{MCAPL_journal}, we therefore provide only a brief overview
here.  It has two main sub-components: the \ail-toolkit for
implementing interpreters for rational agent programming languages and
the \ajpf\ model checker.

\begin{sloppypar}
Interpreters for BDI languages are programmed by instantiating the
\java-based {\em AIL toolkit}~\cite{AIL07:ProMAS}.  An agent system
can be programmed in the normal way for a language but runs in the AIL
interpreter which in turn runs on the Java Pathfinder (\jpf) virtual
machine.  This is a \java\ virtual machine specially designed to
maintain backtrack points and explore, for instance, all possible
thread scheduling options (that can affect the result of the
verification)~\cite{VisserHBPL03}.
\end{sloppypar}

Agent \jpf\ (\ajpf) is a customisation of \jpf\ that is optimised for
AIL-based language interpreters.  Agents programmed in languages that
are implemented using the \ail{}-toolkit can thus be model checked in
\ajpf.  Furthermore if they run in an environment programmed in \java,
then the whole agent system can be model checked.  Common to all
language interpreters implemented using the \ail\ are the \ail-agent
data structures for \emph{beliefs}, \emph{intentions}, \emph{goals},
etc., which are subsequently accessed by the model checker and on
which the modalities of a property specification language are
defined. Since we will be using this \ajpf{} {\em property
  specification language} (\psl) extensively later in this paper, we
here describe its syntax and semantics (from~\cite{MCAPL_journal}).

\subsubsection{Temporal and Modal Logics and the \mcapl\ Property Specification Language}
So far we have mentioned little about the logical description that we
use to describe the properties required of our system. In standard
model checking, the logical formul\ae\ are typically from some variety
of \emph{temporal logic}, such as PLTL (propositional linear temporal
logic). Underlying this particular logic is a model of time based upon
the natural numbers, and formul\ae\ are evaluated at the current
moment in time within this. Thus, some proposition
`$\mathsf{raining}$' might be true at the current moment in time but
false in the next moment. Temporal operators then allow us to move
between moments in time within our formul\ae{}. In PLTL, such
operators are typically of the form `$\always$' (at all future
moments), `$\sometime$' (at some future moment), `$\tlnext$' (at the
next moment), etc. Thus
$$
\sometime \mathsf{raining}\ \imp\ \tlnext \mathsf{get\_umbrella}
$$ 
means that if, at some moment in the future it will be raining, then
at the next moment in time we should get our umbrella. Such a logical
base provides well-understood and unambiguous formalism for describing
the basic system dynamics~\cite{Practical11}. (Note that we do not
define $\sometime$, $\always$, etc, directly but give the semantics of
the more expressive temporal operators $\until$ and $\release$, and
then derive other temporal operators from these as needed.)

However, in order to describe the behaviour of, and requirements
within, rational agents, we often need additional logical
dimensions~\cite{Kurucz07:hbook}. Thus we combine our basic temporal
logic with logical modalities for describing beliefs (that represent
the agent's knowledge about its situation) and motivations (that
represent the agent's reasons for making choices). 

As an example, consider the following formula, a variation of which we
will use later:
$$
\mathbf{B} \: \mathsf{found\_human}\ \imp\ \sometime(\mathbf{B} \:
\mathsf{is\_free}\llor \mathbf{I} \: \mathsf{make\_free})
$$
meaning that if a robot believes it has found a trapped human then
eventually either it believes the human is free or it intends to make
the human free. Notice how this formula combines temporal operators
with the belief operator, $\mathbf{B}$, and the intention operator,
$\mathbf{I}$.

The \ajpf\ property specification language is thus formally defined as follows:

\paragraph{\psl{} Syntax}
The syntax for property formul\ae\ $\varphi$ is as follows, where
$\AILagent$ is an ``agent constant'' referring to a specific agent in
the system, and $f$ is a ground first-order atomic formula:
\begin{equation*}
\begin{array}{rl}
\varphi ::= & \lbelief{\AILagent}{f} \mid \lgoal{\AILagent}{f} \mid
  \lactions{\AILagent}{f} \mid \lintention{\AILagent}{f} \mid
  \lpercept{f}
  \mid \varphi \vee \varphi \mid \neg \varphi
   \mid \varphi \until \varphi \mid \varphi \release \varphi
\end{array}
\end{equation*}
Here, $\lbelief{\AILagent}{f}$ is true if $\AILagent$ believes $f$ to
be true, $\lgoal{\AILagent}{f}$ is true if $\AILagent$ has a goal to
make $f$ true, and so on (with $\lactionsfunc$ representing actions,
$\lintentionfunc$ representing intentions, and $\lperceptfunc$
representing percepts, i.e., properties that are true in the external
environment (e.g., a simulation of the real world) in which the agent
operates).

\paragraph{\psl{} Semantics}
We next examine the specific semantics of our property formul\ae.
Consider a program, $P$, describing a multi-agent system and let
$\MAS$ be the state of the multi-agent system at one point in the run
of $P$.  $\MAS$ is a tuple consisting of the local states of the
individual agents and of the environment. Let $\AILagent \in \MAS$ be
the state of an agent in the $\MAS$ tuple at this point in the program
execution. Then
$$
\MAS \models_{MC} \lbelief{\AILagent}{f} \text{\quad iff\quad }
\AILagent \models f
$$
where $\models$ is logical consequence as implemented by the agent
programming language.  The interpretation of $\lgoal{\AILagent}{f}$ is
given as:
$$
  \MAS \models_{MC} \lgoal{\AILagent}{f} \text{\quad iff\quad } f \in \AILagent_{G}
$$
where $\AILagent_{G}$ is the set of agent goals (as implemented by the
agent programming language). The interpretation of
$\lactions{\AILagent}{f}$ is:
$$\MAS \models_{MC}
\lactions{\AILagent}{f}$$ if, and only if, the last action changing
the environment was action $f$ taken by agent $\AILagent$.  Similarly,
the interpretation of $\lintention{\AILagent}{f}$ is given as:
$$\MAS \models_{MC} \lintention{\AILagent}{f}$$
if, and only if, $f \in \AILagent_{G}$ and there is an {\em intended
  means} for $f$ (in \ail\ this is interpreted as having selected some
plan that can achieve $f$).  Finally, the interpretation of
$\lpercept{f}$ is given as:
$$\MAS
\models_{MC} \lpercept{f}$$ if, and only if, $f$ is a percept that
holds true in the environment.
\medskip

\noindent The other operators in the \ajpf{} property specification
language have standard PLTL semantics~\cite{emerson:90a} and are
implemented as B\"{u}chi Automata as described
in~\cite{Gerth:1995:SOA:645837.670574,Courcoubetis92memory-efficientalgorithms}. Thus,
the classical logic operators are defined by:
$$
\begin{array}{lcl}
  \MAS \models_{MC} \varphi \vee \psi &\text{iff}& \MAS \models_{MC}
  \varphi \text{ or } \MAS \models_{MC} \psi \\
  \MAS \models_{MC} \neg \varphi &\text{iff}& \MAS \not\models_{MC} \varphi.
\end{array}
$$
The temporal formul\ae\ apply to runs of the programs in the \jpf{}
model checker. A run consists of a (possibly infinite) sequence of
program states $\MAS_i$, $i \geq 0$ where $\MAS_0$ is the initial
state of the program (note, however, that for model checking the
number of \emph{different} states in any run is assumed to be
finite). Let $P$ be a multi-agent program, then
$$
\begin{array}{rcl}
  \MAS \models_{MC}\quad \varphi \until \psi 
   &\quad\text{iff}\quad&
    \text{ in all runs of $P$ there exists a state } \MAS_j\\
  &&  \text{ such that } \MAS_i \models_{MC} \varphi \text{ for all } 0 \leq i < j \text{ and } \MAS_j \models_{MC} \psi\\[.5em]
  \MAS \models_{MC}\quad  \varphi \release \psi  
   &\quad\text{iff}\quad&
    \text{ either } \MAS_i \models_{MC} \varphi \text{ for all } i
    \text{ or there exists } \MAS_j \\
  && \text{ such  that } \MAS_i \models_{MC} \varphi
     \text{ for all } i\in\{0,\dots, j\} \text{ and } \MAS_j \models_{MC} \varphi \land \psi
\end{array}
$$
The common temporal operators $\sometime$ (eventually) and $\always$
(always) are, in turn, derivable from $\until$ and $\release$ in the
usual way~\cite{emerson:90a}.

For the future, we hope to incorporate modalities for knowledge (for
describing information evolution) and coalitions (for describing
cooperative systems)\cite{modal:logic:handbook} as well as extending
some or all of these modalities to a full logic.  
Investigating alternative logics for model checking autonomous systems
is also the subject of further work.

\subsection{Abstracting from the Real World}
\label{sec:real_world}
Dealing with the ``real world'' presents a dual challenge to an agent
based programming and analysis system.  Rational agent reasoning, as
supported by BDI languages, is essentially discrete while the real
world is typically not only continuous but also unpredictable.
Similarly, when we wish to analyse a system embedded in the real world,
while we would ideally like to be \emph{certain} about how it will behave,
full/exhaustive formal verification of all aspects of an
autonomous system together with its environment is impossible. Thus,
we cannot say, for certain, whether any course of action our system
takes will definitely achieve/avoid some situation. A typical way to
deal with both these problems has been to use {\em abstraction}.

\paragraph{Abstraction and Decision-Making.}
Dealing with the real world presents a number of problems to the
integration of a control system with a discrete decision-making part;
recall Fig.~\ref{fig:hybrid2}.

Our approach to autonomous decision-making in a hybrid agent system
arises from a careful choice of abstractions that relate the
continuous world with discrete decision states selected by taking into
account all the possible responses of the continuous physical
environment and communications from other agents. Using these
abstractions one can then define basic rules of behaviour, and can
formulate \emph{goals} to keep the system within constraints and to
set both short and long term objectives. This allows us to use the
agent-based approach, where goals, plans, and logical inference are
all captured within the rational agent.

Generating appropriate abstractions to mediate between continuous and
discrete parts of a system is the key to any link between a control
system and a reasoning system.  Abstractions allow concepts to be
translated from the quantitative data derived from sensors (and
necessary to actually run the underlying system) to the qualitative
data needed for reasoning.  For instance a control system may sense
and store precise location coordinates, represented as real numbers,
while the reasoning system may only be interested in whether a vehicle
is within reasonable bounds of its desired position.

We have been exploring an architecture~\cite{DALT10:abstraction} in
which the rational agents are partnered with an \emph{abstraction
  engine} that discretizes the continuous information in an explicit
fashion which we are able to use in verification.  Currently the
abstraction engine is implemented as a distinct rational agent, though
we intend to move to a stream processing model in future.

\paragraph{Abstraction and Formal Verification.}
Abstraction has a long history in the formal modelling of ``real
world'' phenomena for verification~\cite{Clarke:1994:Abs}.
Abstractions for model checking hybrid systems have been extensively
studied~\cite{AlurCHHHNOSY95,HenzingerHW97,AbstrHybSys08,Book:VerificationHybridSys09}.
Model checking requires the continuous search space to be divided into
a finite set of regions which can be examined, in order for the model
checker to exhaustively trace a finite set of paths through these
regions.  This provides a simpler representation of the world which
(hopefully) preserves its key features.

Since the ``real world'' is inherently complex, often involving
control systems, physical processes and human interactions, even
defining appropriate abstractions is difficult. Often we either take a
very \emph{coarse} abstraction, which risks being very far away from a
real system, or a very \emph{detailed} abstraction leading to complex
structures such as stochastic, hybrid automata (which are, themselves,
often very hard to deal with~\cite{Manuela12:book}). Neither of these
are entirely satisfactory, and neither guarantees completely accurate
analysis.  Our approach opts for a coarse abstraction which
over-approximates the real world.  Where we have an explicit
abstraction engine, as mentioned above, we use this to form the coarse
abstraction (see the examples in sections~\ref{sec:case_study} and
\ref{sec:adaptive}).  In many ways this is similar to the approach
taken by \phaver\ but is not based on a decomposition of the system
into parallel hybrid automata; we are also less interested in
verifying the whole system but are primarily concerned with verifying
the decision-making algorithms.
\medskip

\noindent In this section we have examined the history of agent-based
control of hybrid systems and the separate strands of research into
the verification in hybrid systems and verification of agent programs.  
We next begin to tackle a series of example systems, defining the
controlling agent(s), describing the environmental assumptions and
carrying out formal verification.

\section{Scenario:~Urban Search and Rescue}
\label{sec:scen2}
Consider a (simplified) example based on the \emph{RoboCup Rescue}, or
``urban search and rescue'', scenario~\cite{RoboCupRescue01}. This is
a widely used test scenario for autonomous mobile robots, providing
not only a variety of environmental abstractions but also the
potential for both individual and cooperative behaviours. A natural
disaster (e.g., an earthquake) has caused buildings to
collapse. Autonomous robots are searching for survivors buried in the
rubble. There may be many of these robots, each with sensors for
detecting buried humans. Let us model this scenario on a simple
grid. A robot can move around spaces on the grid and a human is placed
randomly on the grid. The robot's sensors are such that a robot
`detects' a human if the human is in the same grid
square\footnote{Relevant code for all the examples in this paper can be found in the
  examples package of the \mcapl\ Project, available at sourceforge:
  \url{http://mcapl.sourceforge.net} or can be supplied on request by the first author.}.

\subsection{Implementation}
Listing~\ref{code:simple} shows part of the agent code for a
simple search robot written using the BDI-style agent programming
language, \gwendolen~\cite{dennis08:_gwend}. The robot has the goal of
reaching a state where it can leave the area. It can only achieve this
goal if either it believes it has found the human or believes that the
area is actually empty.

\begin{lstlisting}[float,caption=Simple Rescue Robot,basicstyle=\sffamily,style=easslisting,language=Gwendolen,label=code:simple]
Rules
B area_empty :- lnot(B square(X,Y) & lnotB empty(X,Y));
B unchecked(X,Y) :- B square(X,Y) & lnotB at(X,Y) & 
                      lnotB empty(X,Y) & lnotB human(X,Y);

Plans
+!leave : { lnotB at(X1,Y1), B unchecked(X,Y) } <- +at(X,Y), move_to(X,Y);
+!leave : { B at(X,Y), lnotB human(X, Y), lnotB area_empty, B unchecked(X1,Y1) } <-
     +empty(X,Y),
     -at(X,Y),
     +at(X1,Y1),
     move_to(X1,Y1);
+!leave : { B at(X,Y), lnotB human(X, Y), lnotB area_empty, lnotB unchecked(X1,Y1) } <-
     +empty(X,Y),
     -at(X,Y);
+!leave : { B at(X,Y), B human(X, Y) } <- +found;
+!leave : { B area_empty } <- +leave;

+found : { B at(X,Y) } <- send(lifter, B human(X,Y)), 
                          +sent(lifter, human(X,Y)), 
                          +leave;
\end{lstlisting}

\paragraph{Syntax.} \gwendolen{} uses many syntactic conventions from
BDI agent languages: {\lstinline{+!g}} indicates the addition of the
goal {\lstinline{g}}; {\lstinline{+b}} indicates the addition of the
belief {\lstinline{b}}; while {\lstinline{-b}} indicates the removal
of the belief. Plans then consist of three parts, with the pattern

\begin{small}
\begin{center}
\lstinline{trigger : guard <- body}.
\end{center}
\end{small}

\noindent The `{\lstinline{trigger}}' is typically the addition of a
goal or a belief (beliefs may be acquired thanks to the operation of
perception and as a result of internal deliberation); the
`{\lstinline{guard}}' states conditions about the agent's beliefs
(and, potentially, goals) which must be true before the plan can
become active; and the `{\lstinline{body}}' is a stack of `deeds' the
agent performs in order to execute the plan. These deeds typically
involve the addition and deletion of goals and beliefs as well as {\em
  actions} (e.g., {\lstinline{move_to(X1,Y1)}}) which refer to code
that is delegated to non-rational parts of the systems (in this case,
motor systems). In the above, \textsc{Prolog} conventions are used,
and so capitalised names inside terms indicate free variables which
are instantiated by unification (typically against the agent's
beliefs).  Programs may also perform deductive reasoning on their
atomic beliefs as described in their {\em belief rules}.  For instance

\begin{small}
\begin{center}
\lstinline{B area_empty :- lnot(B square(X,Y) & lnotB empty(X,Y))}
\end{center}
\end{small}

\noindent indicates that the agent believes the whole area is empty if
it is not the case (i.e., `{\lstinline{lnot}}') that it believes
there is some square that it does not believe is empty (the ``closed
world assumption'' is used to deduce that the agent does not believe
something).

In Listing~\ref{code:simple}, the goal, {\lstinline{+!leave}},
represents an {\em achievement goal}~\cite{RiemsdijkDM09} meaning that
the agent must continue attempting the plans associated with the goal
(if they are applicable) until it has acquired the belief
`{\lstinline{leave}}' (i.e., it wishes to achieve a state of the world
in which it believes it can leave the area).  Note that, in the above
program, the agent cannot directly deduce ``{\lstinline{B human}}'',
i.e., there is a human at the current position, as it requires its
sensors to tell it if it can \emph{see} a human.  The underlying
language interpreter regularly polls the sensors for belief changes
associated with perception.

In Listing~\ref{code:simple} the agent picks an unchecked square
in the grid to explore.  It continues to do this until either its
sensors tell it that it can see a human or until it believes the area
is empty.

\subsection{Verification}
Now, we wish to verify the robot's reasoning is correct, {\em
  independent} either of the modelling of other parts of the system or
any detail of its environment.  So, we might wish to verify that
\begin{quote}
{\em if the searching robot, $s$, believes it can leave the
area, then it either believes the human is found or it believes the
area is empty.}
\end{quote}
\begin{equation}
\label{thm:leave}
\always (\mathbf{B}_s\, \mathsf{can\_leave} \imp (\mathbf{B}_s\,
\mathsf{found}\, \lor\, \mathbf{B}_s\, \mathsf{area\_empty}))
\end{equation}
As noted in Section~\ref{sec:psl}, we used the \ajpf\ model checker
for this verification.  

An abstract model for the incoming perceptions must be provided,
otherwise we would only be able to check the robot's behaviour when it
does not believe it can see a human (since that belief arises from
perception alone). Whenever the agent requests some perceptions we
supply, randomly, all relevant inputs to the robot. In this case it
either sees a human or does not see a human.
The choice of which perceptions to supply to the agent and the times
at which these perceptions may change obviously represent an {\em
  abstract} model of the real world.  However we believe that these
are decisions that can be made explicit, avoiding the need for someone
to inspect the code of the model in order to determine precisely what
has been proved.  Moreover it is a model that is extremely
conservative in its assumptions about the real world.

Using this model of the environment we were successfully able to
verify \eqref{thm:leave} automatically in \ajpf.  Similarly we also
verified other parts of the agent reasoning for instance, 
\begin{quote}
{\em if the searching robot, s, 
never believes it sees the human then that means that eventually
it will believe it has visited all squares in the grid.}
\end{quote}
\begin{equation}
\label{eq:every_square}
\always\lnot \lbelief{s}{\mathsf{human}}\ \Rightarrow\ \forall\
\mathsf{square(X,Y)}\in \mathsf{Grid}.\ \sometime \lbelief{s}{\mathsf{at(X,Y)}}
\end{equation}
NB. the \ajpf\ property specification language is propositional and so
the use of universal ($\forall$) quantifiers in these examples is just
a shorthand for an enumeration over all possible values.  In this case
we used a $3\times 3$ grid so the actual property checked was
$$
\always\left(\begin{array}{c}
\lnot \lbelief{s}{\mathsf{human}}\ \Rightarrow\ 
  \left(\begin{array}{l}
  \sometime \lbelief{s}{\mathsf{at(0, 0)}} \,\wedge\, 
  \sometime \lbelief{s}{\mathsf{at(0, 1)}} \,\wedge\, 
  \sometime \lbelief{s}{\mathsf{at(0, 2)}} \,\wedge\, \\
  \sometime \lbelief{s}{\mathsf{at(1, 0)}} \,\wedge\, 
  \sometime \lbelief{s}{\mathsf{at(1, 1)}} \,\wedge\, 
  \sometime \lbelief{s}{\mathsf{at(1, 2)}} \,\wedge\, \\
  \sometime \lbelief{s}{\mathsf{at(2, 0)}} \,\wedge\, 
  \sometime \lbelief{s}{\mathsf{at(2, 1)}} \,\wedge\, 
  \sometime \lbelief{s}{\mathsf{at(2, 2)}}
  \end{array}\right)
\end{array}\right)
$$
Use of quantifiers in other \ajpf\ property specification language
expressions throughout the paper should be understood in a similar
way.

\subsubsection{Deduction using Environmental Assumptions}
There is a great deal we can formally verify about an agent's
behaviour {\em without} reference to the agent's environment.  However
there are still things we might reasonably wish to prove given simple
assumptions about the behaviour of the real world.  Thus, once we have
shown $\mathit{System}\models\varphi$, for some relevant property
$\varphi$, then we can use the fact that the environment satisfies
$\psi$ to establish $\mathit{System}\models(\psi\land\varphi)$ and
hence (if $(\psi\land\varphi)\Rightarrow\xi$) that
$\mathit{System}\models\xi$.
For instance, we might want to assume that the robot's sensors
accurately detect the human, and that its motor control operates
correctly.  If we know these facts then we can prove a stronger
property that the agent will \emph{actually} find the human as well as
believing it has found the human.

Let us define `$\mathsf{found\_human}$' as the property that 
\begin{quote}
{\em the robot
and the human are in the same grid square {\em and} the robot believes
it can see the human.}
\end{quote}
\begin{equation}
\begin{array}{c}
\mathsf{found\_human}\quad \equiv\quad 
\lbelief{s}{\mathsf{human}} \land
 (\exists\, \mathsf{square(X,Y)}\in
\mathsf{Grid}.\ \mathsf{at(human,X,Y)} \land \mathsf{at(robot,X,Y)})
\end{array}
\label{eq:found_human}
\end{equation}

\noindent We can characterise $\mathsf{correct\_sensors}$ as:
\begin{quote}
{\em the robot believes it can see a human if, and only if, it is in
the same square as the human.}
\end{quote}
\begin{equation}
\begin{array}{c}
  \!\!\mathsf{correct\_sensors}~~\equiv~~(\lbelief{s}{\mathsf{human}}\ \iff\ (\exists\, \mathsf{square(X,Y)}\in
  \mathsf{Grid}.\ \mathsf{at(human,X,Y)} \land \mathsf{at(robot,X,Y)}))
\end{array}
\end{equation}
Similarly, we need to state, as an assumption, that the robot's
motors are working correctly.
\begin{quote} {\em the robot believes it has reached a grid square if,
    and only if, it is actually in that grid square.}
\end{quote}
\begin{equation}
\begin{array}{c}
\mathsf{correct\_movement}\quad \equiv\quad
\forall\, \mathsf{square(X,Y)}\in
\mathsf{Grid}.\  \always (\lbelief{s}{\mathsf{at(X,Y)}}\, \iff\, \mathsf{at(robot,X,Y)})
\end{array}
\end{equation}

\noindent Given this framework we might
want to show that
\begin{quote}
 {\em if the robot's sensors and motors are working correctly and if a
   human is stationary on the grid then eventually the robot will find
   the human.}
\end{quote}
\begin{equation}
\label{eq:deductive}
\begin{array}{c}
  \left[\begin{array}{c}
      \always \mathsf{correct\_sensors}\, \land\, \always 
      \mathsf{correct\_movement}\, \land\\
      \exists\,  \mathsf{square(X,Y)}\in\ \mathsf{Grid}.\ \always
      \mathsf{at(human,X,Y)}
\end{array}\right]\ \imp\ \sometime \mathsf{found\_human}
\end{array}
\end{equation}
We have already verified the agent's internal reasoning by model
checking to give us the property in~\eqref{eq:every_square} which
tells us that either the robot believes it sees the human or it visits
every square.  From this~\eqref{eq:deductive} can easily be
proved. This can be done by hand, or by using a suitable
temporal/modal logic prover~\cite{Practical11} - for instance we were
easily able to prove this theorem on a three by three grid
automatically in the online Logics
Workbench~\cite{DBLP:journals/aicom/HeuerdingJSS96} (see
Appendix~\ref{app:lwb} for the input commands).  \medskip

\noindent A key aspect here is that this `proof' that the robot will
find the human \emph{only} works if the human does not move and,
indeed, we have had to state that explicitly (via $\always
\mathsf{at(human,X,Y)}$)!  Without this assumption the proof fails
since there will be no way to prove that, eventually, the robot and
human will be in the same square at the same time.

\subsubsection{Multi-Agent Systems}
\label{sec:eass:rw}
In more sophisticated scenarios we not only want to check properties
of single agents but of groups of agents working together.  Thus,
imagine that we now have another robot, capable of `lifting'
rubble. The aim is for the two robots to work as a team: the
`searching' robot, `$s$', will find the human, then the `lifting'
robot, `$l$', will come and remove the rubble.  We will refer to the
beliefs of the lifting robot as $\mathbf{B}_l$.
Ideally, if these two work together as expected
then we would like to show that eventually the lifter believes the
human is free:
\begin{equation}
\sometime\mathbf{B}_l \mathsf{free(human)}
\end{equation}
However, this depends on several things, for example that any
communication between the robots will actually succeed. We can adopt
our previous approach and analyse each robot independently based on
random perceptions and, in this case, messages being received from the
environment.  So we can establish that (we have slightly simplified
the properties for presentational purposes):
\begin{quote}
{\em the searcher will send a message to the lifter if it finds a
human.}
\end{quote}
\begin{equation}
\always (\mathbf{B}_s \mathsf{found} \imp
\sometime \mathbf{B}_s \mathsf{sent(lifter, human(SomeX, SomeY))})
\label{eq:searcher_prop}
\end{equation}
and
\begin{quote}
{\em the lifter will free the human if it receives such a message}
\end{quote}
\begin{equation}
\always(\mathbf{B}_l
\mathsf{rec(searcher, human(X, Y))}) \imp
\sometime \mathbf{B}_l \mathsf{free(human)})
\label{eq:lifter_prop}
\end{equation}

\noindent We can also express the assumption that messages sent by the searcher
will always be received by the lifter and then use the model checker
to prove properties of the combined system, e.g:
\begin{quote}
{\em if messages sent by the searcher are always received by the lifter
then if the searcher believes it has found a human, eventually the
lifter will believe the human is free.}
\end{quote}
\begin{equation}
\begin{array}{c}
\always (\lbelief{s}{\mathsf{sent(lifter, human(X, Y))}}
\imp
\sometime \lbelief{l}{\mathsf{rec(searcher, human(X, Y))}})\\
 \imp\quad \always (\mathbf{B}_s \mathsf{found} \imp  \sometime\mathbf{B}_l \mathsf{free(human)})
\end{array}
\label{eq:combined}
\end{equation}

\noindent Potentially, using reasoning such as this, two autonomous
robots could believe that they will together achieve a required
situation given some joint beliefs about the environment.  However if
we reason about each agent separately to deduce
properties~\eqref{eq:searcher_prop} and~\eqref{eq:lifter_prop} we
reduce both the size of the automaton to be checked and the size of
the search space.  We can then combine these component properties with
an appropriate statement about communication, i.e.
$$\always (\mathbf{B}_s
\mathsf{sent(lifter, human(X, Y))}
\imp \
\sometime \lbelief{l}{\mathsf{rec(searcher, human(X, Y))}})$$
in order to reach the conclusion of~\eqref{eq:combined}
deductively\footnote{Logics Workbench Commands, again in
  Appendix~\ref{app:lwb}.}.  This demonstrates one of the advantages of
a compositional approach -- namely that the complexity of model
checking tasks can be kept to a minimum.

\subsubsection{Goals and Intentions}
We have been verifying the beliefs agents acquire about their
environment \emph{in lieu} of verifying actual \emph{facts} about the
environment.  However, we are also interested in verifying the choices
that agents make. Suppose that our lifting agent does \emph{not}
deduce that the human is free (because it has moved some rubble), but
continues to lift rubble out of the way until its sensors tell it the
area is clear (see Listing~\ref{code:intention}).
\begin{lstlisting}[float,caption=Lifting Robot,basicstyle=\sffamily,style=easslisting,language=Gwendolen,label=code:intention]
+received(searcher, Msg): { lnotB Msg } <- +Msg;

+human(X,Y): { True } <- +!free(human);

+!free(human):{ B human(X,Y), lnotB at(X,Y), lnotB clear } <- move_to(X,Y), +at(X,Y);
+!free(human):{ B human(X,Y) , B at(X,Y), lnotB clear } <- lift(rubble);
+!free(human):{ B human(X,Y), B at(X,Y),  B clear, lnotB have(human) } <- 
    lift(human), 
    +have(human);
+!free(human):{ B have(human) } <- +free(human);
\end{lstlisting}
We cannot verify that the robot will eventually believe the
human is free since we can not be sure that it will ever believe that
the human is actually clear of rubble. However, we can establish (and
have verified) that
\begin{quote}
{\em  if the lifting agent believes the human to be at position $\mathsf{(X,Y)}$
then eventually it will form an intention to free the human.}
\end{quote}
\begin{equation}
\mathbf{B}_l \mathsf{human(X,Y)} \imp
\sometime (\mathbf{I}_l \mathsf{free(human)} \lor \mathbf{B}_l \mathsf{free(human)})
\end{equation}

\noindent As above, we can derive further properties under assumptions
about the way perception behaves:
\begin{quote}
{\em assuming that, whenever the lifter forms an intention to free the
human it will eventually believe the rubble is clear, then 
receipt of a message from the searcher will eventually
result in the lifter believing the human is free.}
\end{quote}
\begin{equation}
\always (\mathbf{I}_l \mathsf{free(human)}
\imp \sometime \mathbf{B}_l \mathsf{clear}) \quad\imp \quad
(\mathbf{B}_l \mathsf{rec(searcher, found)} \imp \sometime
\mathbf{B}_l \mathsf{free(human)})
\end{equation}

\noindent While much simplification has occurred here, it is clear how
we can carry out compositional verification, mixing agent model
checking and temporal/modal proof, and how the environmental
abstractions we use can be refined in many ways to provide
increasingly refined abstractions of the ``real world''. Crucially,
however, we can assess the choices the agent makes based on its
beliefs about its environment and not what actually happens in its
environment~\cite{FDW12:magazine}.  

\section{Scenario:~Rational Hybrid Agents for Autonomous Satellites}
\label{sec:eass:arch}
The previous example involved simple code developed to illustrate
our methodology.  We now turn to look at code developed as part of a
project to investigate agent based control of satellite systems~\cite{lincoln13:_auton_aster_explor_ration_agent}.  The
code was not initially developed with verification in mind.

%
Traditionally, a satellite is a large and very expensive piece of
equipment, tightly controlled by a ground team with little scope for
autonomy. Recently, however, the space industry has sought to abandon
large monolithic platforms in favour of multiple, smaller, more
autonomous, satellites working in teams to accomplish the task of a
larger vehicle through distributed methods.

The nature of these satellite systems, having a genuine need for
co-operation and autonomy, mission critical aspects and interaction
with the real world in an environment that is, in many respects,
simpler than a terrestrial one, makes them a good test-bed for our
approach to analysing autonomous systems.

\subsection{System Architecture}

We have built a hybrid system embedding existing technology
for generating feedback controllers and configuring satellite systems
within a decision-making part based upon a high-level agent
program.
%
The decision-making relies on {\em discrete} information (e.g., ``a
thruster is broken'') while system control tends to rely on {\em
  continuous} information (e.g., ``thruster fuel pressure is 65.3'').
Thus, it is vital to be able to \emph{abstract} from the dynamic
system properties and provide discrete abstractions for use by the
agent program. It is for this reason that, as mentioned earlier, we have an explicit
\emph{abstraction layer} within our architecture that translates
between the two information styles as data flows around the system.

\begin{figure}[htbp]
\begin{center}
\includegraphics[width=.817625\textwidth]{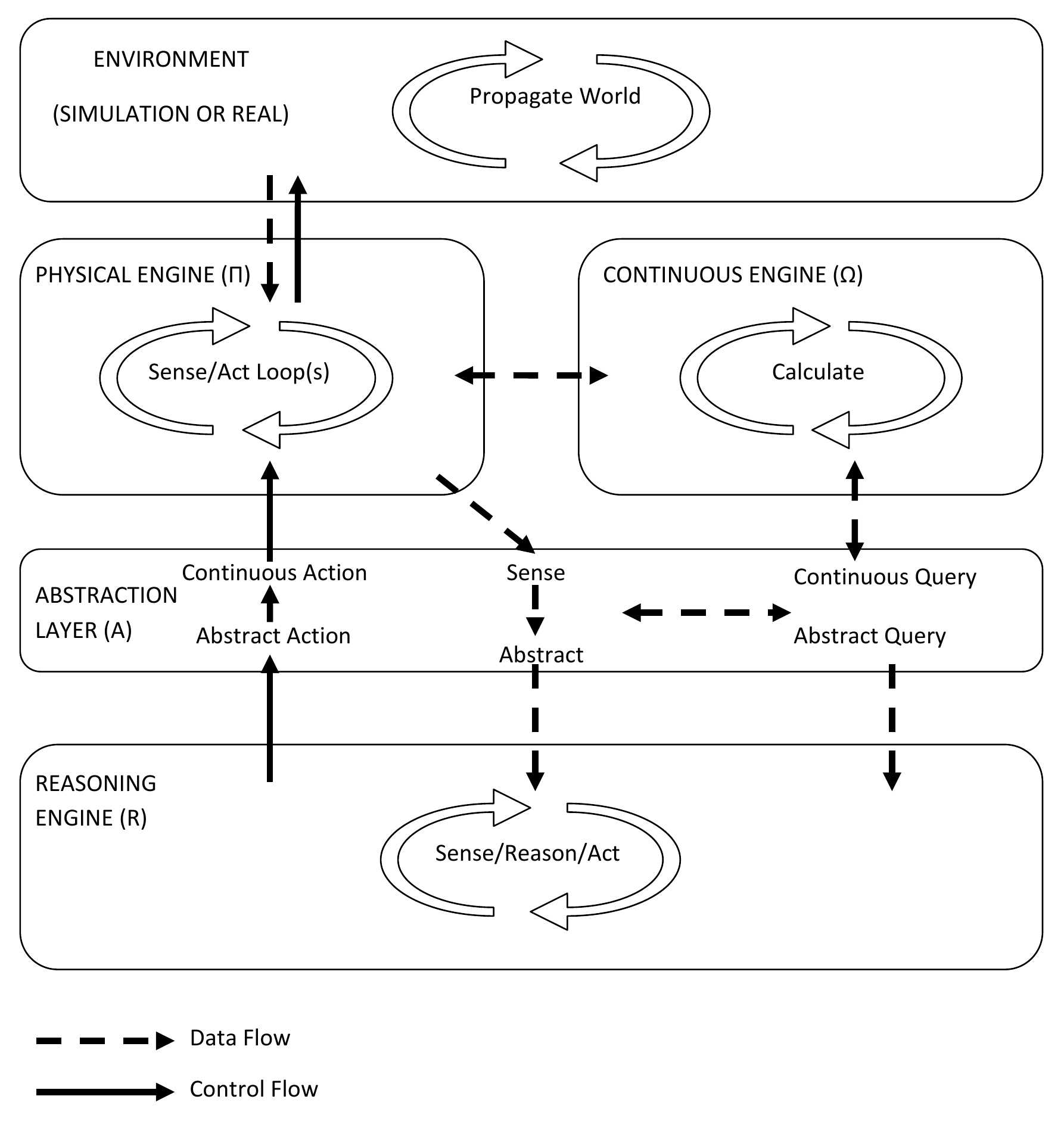}
\end{center}
\caption{Hybrid Agent Architecture.  Real time control of the satellite is
governed by a traditional feedback
controller 
drawing its sensory input from the environment.  This forms a {\em
  Physical Engine} ($\Pi$).  This engine communicates with an agent
architecture consisting of an {\em Abstraction Engine} ($A$) that
filters and discretizes information.
To do this $A$ may use a {\em Continuous Engine} ($\Omega$) to make
calculations.  Finally, the {\em
  Rational Engine} ($R$) contains a ``Sense-Reason-Act'' loop captured
as a rational agent. Actions involve either calls to the Continuous
Engine, $\Omega$, to calculate new controllers (for instance) or
instructions to the Physical Engine, $\Pi$, to change these controllers.  These instructions are passed through the Abstraction
Engine, $A$, for reification.}
\label{fig:abs_arch}
\end{figure}

Fig.~\ref{fig:abs_arch} shows the full architecture for our
system~\cite{DFLLV10:IEEE}.  
$R$ is a traditional BDI system dealing with
discrete information, $\Pi$ and $\Omega$ are traditional control
systems, typically generated by \matlab{}/\simulink{}, while $A$
provides the vital ``glue'' between all these
parts~\cite{DALT10:abstraction}.  We
will not discuss the implementation of $\Pi$ and $\Omega$ in any
detail; our concern is primarily with $R$ and its interaction with
$A$.

The agent programming language used to implement $R$ (again,
\gwendolen) encourages an engineer to express decisions in terms of
the beliefs an agent has, what it wants to achieve, and how it will
cope with any unusual events. This reduces code size so an engineer
need not explicitly describe how the satellite should behave in each
possible configuration of the system, but can instead focus on
describing the factors relevant to particular
decisions~\cite{DFLLV10:iSairas}. The key aspect of
\emph{deliberation} within agent programs allows the decision-making
part to adapt intelligently to changing dynamic situations, changing
priorities, and uncertain sensors.

\subsubsection{Semantics of Interaction}
\label{sec:semantics}

Our hybrid architecture consists of a Physical Engine
($\Pi$), a Continuous Engine ($\Omega$), an Abstraction Engine ($A$)
and a Reasoning Engine ($R$). We developed a semantics for interaction
between these which operates via shared sets of which, for our
purposes, the most interesting is the set of {\em shared beliefs},
$\Sigma$, used by both $A$ and $R$ to describe their shared, discrete,
understanding of the world.  $R$ may also
make specific requests for calculations to be performed by $\Omega$ or
actions to be performed by $\Pi$.  These requests are sent via $A$.
Among other things, $A$ takes any (potentially continuous) results
returned by $\Omega$ and discretizes these as a shared belief for
$R$. Overall, the role of $A$ is to interface between $R$ and the
other engines.  This helps avoid data overload by replacing large
`chunks' of incoming data with key discrete predicates~\cite{DALT10:abstraction}.

In this semantics all perceptions that arrive via sensors from the
real world are filtered through $A$, which converts them into discrete
shared beliefs, $\Sigma$. Therefore, from a model checking
perspective, if we are interested in how the external world can affect
the internal beliefs of the Reasoning Engine, $R$, then we are
primarily interested in the possible compositions of $\Sigma$.

\subsubsection{System Implementation}
This architecture and interaction semantics have been implemented
within a simulation environment, where the Physical and Continuous
engines ($\Pi$ and $\Omega$) are implemented in \matlab{}, while $A$
and $R$ are written in a customised variant of the \gwendolen{}
programming language~\cite{dennis08:_gwend}.  This extended language
includes constructs for explicitly calling $\Pi$ and $\Omega$ and a
\java\ environment interface that supports such calls.  This \java{}
environment handles the shared data sets, in particular the shared
beliefs which are used by $A$ and $R$ and also controls communication
with \matlab{} via sockets.


The architecture has been deployed on multiple satellites within a
satellite hardware test facility as well as within a number of
simulated environments. A range of satellite scenarios have been
devised and tested, involving assuming and maintaining various
formations, fault monitoring and recovery and collaborative surveying
work in environments such as geostationary and low Earth orbits and
among semi-charted asteroid fields.  The system and scenarios are
described more fully
in~\cite{lincoln13:_auton_aster_explor_ration_agent}.

The focus of this example is upon the verification that took place {\em
  after} this initial implementation and testing.  Given the verification was of the rational
engine {\em alone}, it was obviously important that the behaviour of
the whole system be tested separately but this paper does not examine these aspects.

\subsection{Adapting the System for Model Checking}
Our principal interest here is in the verification of the discrete
reasoning parts of the system. These are represented by $R$ and so we
want to abstract away $A$, $\Omega$ and $\Pi$, yet do so in a coherent
fashion.  Since all communication with $\Pi$ and $\Omega$ to and from
$R$ occurs via $A$, we can ignore $\Omega$ and $\Pi$ entirely and
just focus on $A$ and $R$.


The architecture proved particularly conducive to the verification
methodology proposed.  Since programmers were encouraged to explicitly
program up the abstractions to be used with the Abstraction Engine,
$A$, it became possible to ``read off'' from the code for an
abstraction engine all the shared beliefs that could be asserted in
response to changing perceptions.  This in turn allows us to
pre-determine the possible configurations of $\Sigma$, the set of
shared beliefs.  Similarly we were able to analyse the messages that
agents sent to determine which messages might be received by other
agents in the system.  Since the only external input to the reasoning
engine was from shared beliefs and messages it was easy to clearly
define the set of inputs needed for verification.

We implemented a specific {\em verification environment} that observed
the Reasoning Engine's interface requirements for the hybrid system.
This consisted entirely of asserting
and retracting these shared beliefs and messages.  Each time an agent took an
action in the environment a new set of shared beliefs and a new set of messages were generated {\em at random}.  Each time the agent requested the shared beliefs
a random shared belief set was sent to it (similarly with messages).  During model checking the calls to random number generation caused the model checker branch and to explore all possible 
outcomes that could be generated.  In order to
limit the search space we take the (simplifying) view that reasoning
happens instantaneously, while action has a duration.  Therefore the
only times at which the system randomly changes the perceptions/shared
beliefs and messages available to the reasoning engine are when the reasoning engine
takes some form of action (i.e., a request for a calculation from
$\Omega$).  The model checker will then explore {\em all possible}
combinations of shared beliefs and messages that might be available at that point,
modelling essentially both the times an action results in the expected
outcome and those when it does not.




\noindent It is important to again emphasize that those aspects of the
system relating to the Abstraction, Continuous and Physical engines
were not required for model checking the system.

\subsection{Example: Autonomous Satellites in Low Earth Orbit}
\label{sec:case_study}

A low Earth orbit (LEO) is an orbit with an altitude ranging
between that of the Earth's upper atmosphere, at approximately
$250$km, and an upper bound of $2000$km; the orbit may be inclined to
the equator and may or may not be elliptical. 
LEOs are used
predominantly by Earth observation missions that require high
resolution imaging, including weather, Earth resource and military
satellites. 

LEO based satellites travel at high speed, completing an orbit within
approximately $90$ minutes. 
Orbiting at such great speeds presents a
secondary issue concerning the control and monitoring of LEO
satellites: ground station visibility is restricted to between $5$ and
$15$ minutes per overhead passage of a ground station. Whilst multiple
ground stations, or space based relay satellites orbiting at higher
altitudes, may be switched between to enable greater communication
periods, the growth in infrastructure is a clear disadvantage.  As a result there is a need to increase the autonomous control of such systems.


\subsubsection{Scenario}

We developed a model
of satellite formations in low Earth orbit. The Continuous Engine,
$\Omega$, was programmed to supply plans for moving to particular
locations in a formation.  The algorithm used to generate such plans
was based on that
in~\cite{lincoln06:_compon_vision_assis_const_auton}. Controls were
made available in the Physical Engine, $\Pi$, which could send a
particular named \emph{activation plan} (i.e., one calculated by
$\Omega$) to the feedback controller and control the valves that
determined which fuel line was being utilised by each thruster.

The satellite was assumed holonomic in control, and provided with
thrusters in three body axes (X, Y and Z) each of which contained two
fuel lines.
This enabled the agent to switch fuel line in the event of a rupture
(detectable by a drop in fuel pressure).

In the simple case examined here, the satellites were expected to move
to pre-generated locations in a formation, correcting for a single
fuel line breakage, if it occurred.

\subsubsection{Implementation and Testing}

The code for the abstraction and rational engines for these examples
can be found in Appendix~\ref{leo} together with a detailed
description of its functionality.

The software was developed and tested both in simulation, using a
\simulink{} implementation of $\Pi$ and $\Omega$, and on a physical
satellite simulation environment developed initially at the University
of Southampton and then moved to the University of Sheffield.  In
simulation a ``gremlin'' agent was introduced that could, at specified
points, insert hardware failures into the satellite system.  In
physical testing it was possible to simulate failures in ``path
following'' by physically moving the satellites from their chosen
course. Having produced a working system the task was then to formally
analyse it.

\begin{figure}[htbp]
\begin{center}
\includegraphics[width=0.9\textwidth]{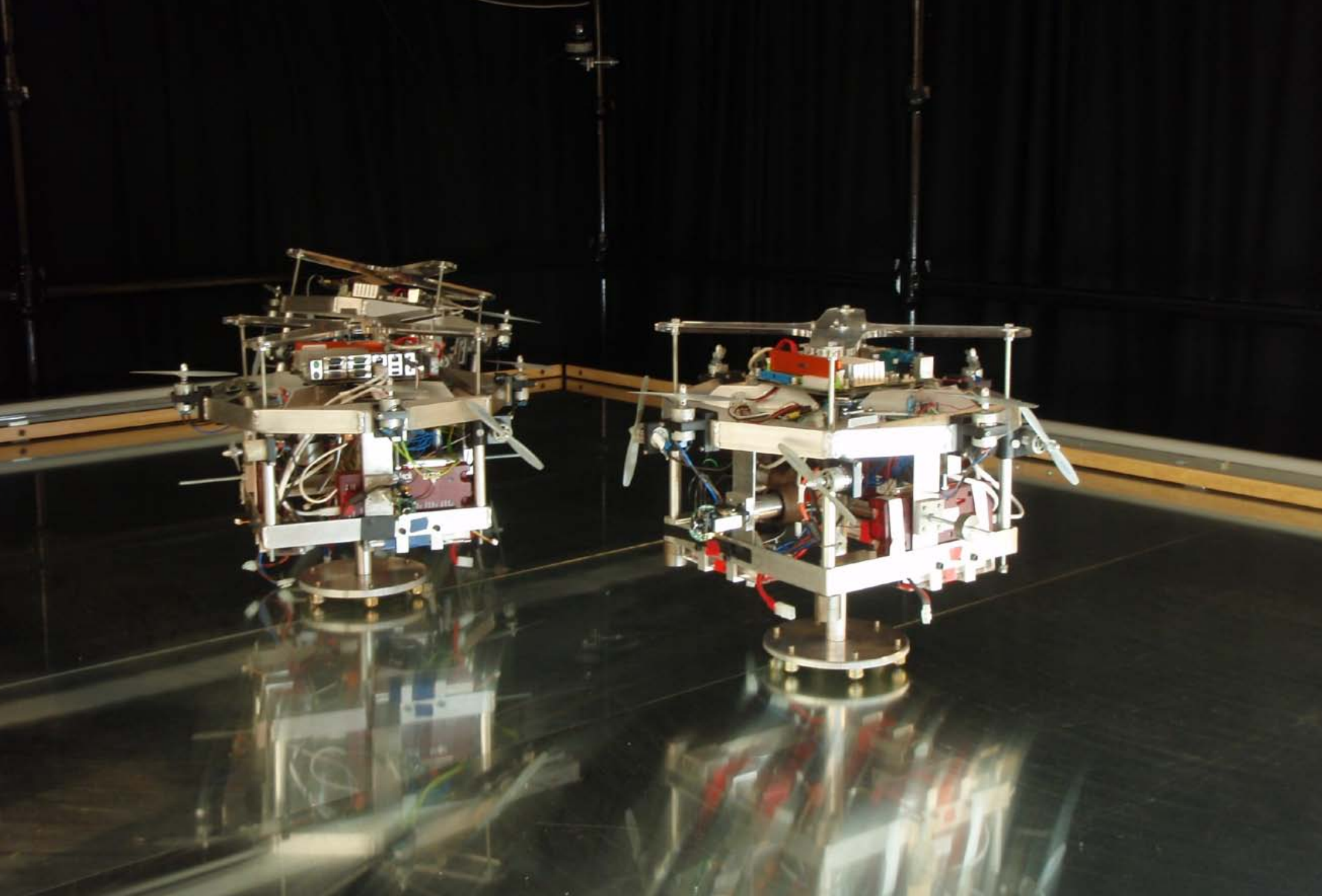}
\end{center}
\caption{The Satellite Test Facility used for Demonstrations of
  Autonomous Agent Implementations.}
\end{figure}

\subsubsection{Analysis of a Single Satellite Agent}
We wish to verify the satellite's Reasoning Engine is correct, {\em
  independent} of the modelling of other parts of the system and any
detail of its environment.

Inspection of the Abstraction Engine code revealed that the shared
beliefs that needed to be were, \lstinline{close_to(Pos)} (meaning
that the satellite is close to position, \lstinline{Pos}),
\lstinline{broken(X)} (meaning thruster \lstinline{X} is broken),
\lstinline{thruster_bank_line(X, B, L)} (meaning thruster
\lstinline{X}, is currently in bank \lstinline{B} and using fuel line
\lstinline{L}) and \lstinline{get_close_to(Pos, P)} (meaning that
\lstinline{P} is a plan for moving to position, \lstinline{P}).
All of these have parameters.  We knew from the physical
satellite set-up that the thruster ``name'', X, could be
{\lstinline{x}}, {\lstinline{y}} or {\lstinline{z}}
and its bank, B (not used in these examples), and line, L, could be
{\lstinline{1}} or {\lstinline{2}} respectively, depending
upon the formation under consideration, $Pos$, could be
{\lstinline{none,}} {\lstinline{right}},
{\lstinline{middle}}, {\lstinline{left}},
{\lstinline{topright}}, {\lstinline{topleft}},
{\lstinline{bottomright}} or {\lstinline{bottomleft}}.
{\lstinline{get_close_to}} also had the name of a plan, $P$, as
a parameter.  This name was generated and supplied by the Continuous
Engine.  This was more of a challenge since the name of the plan would
be derived algorithmically inside the Continuous Engine.  However the
code, itself, never analysed this name simply passing it on to the
Physical Engine and so a place-holder name, {\lstinline{plan}},
was used.

Unfortunately, even with relatively few perceptions the combinatorial
explosion associated with exploring all possibilities gets very
large.  For most experiments it is necessary therefore to limit the
number of possibilities (we discuss this further in Section~\ref{sec:concl}).  However the analysis does allow us to state
clearly which inputs are being considered, as is shown in
tables~\ref{results:single_nothruster}, \ref{results:single}, \ref{results:multi_nofailure_line},
and \ref{results:multi_change}.

The choice of parameters for shared beliefs and the times
at which these shared beliefs might change obviously represent an {\em
  abstract} model of the real world.  However we believe that these
are decisions that can be made explicit, avoiding the need for someone
to inspect the code of the model in order to determine precisely what
has been proved.  Moreover it is a model that is extremely
conservative in its assumptions about the real world.

%
In Section~\ref{sec:eass:rw} we modelled reliable communication via a
hypothesis to our theorem.  We will use the same technique to capture
hypotheses about the workings of the Continuous and Physical engines.
Thus, potentially, once we have shown $\mathit{R}\models (\psi
\rightarrow \varphi)$ then we can use the fact that, for instance
$\Pi$ satisfies $\psi$ to establish $\{\mathit{R},
\Pi\} \models \varphi$

As a result of this analysis a number of additional plans were
introduced into the agent code to handle edge conditions where
information was changing particularly rapidly.  The code shown in
Appendix~\ref{leo} is the final code. The plans on line 50 and lines
56-58 in the Rational Engine in Appendix~\ref{leo:satellite} were
introduced as a result of model-checking to cover cases where, for
instance, an abort has already occurred but information arrives that a
thruster is newly broken, or that a thruster had been successfully
fixed but is now reporting it is broken again.  This demonstrates the
ability of verification to locate errors that were not found by
simulation or testing.

\paragraph{Agent Operation without Thruster Failure or Outside Communication}
We first chose to investigate the operation of the agent in a
situation where its thrusters do not fail and when no messages were
received from outside.  Therefore it started out with the goal of
assuming its position in a formation, and that the position it was to
assume was in the {\lstinline{middle}} of a line. 
This was achieved by preventing the environment from
sending the perception {\lstinline{broken(X)}}.  The environment
delivered no messages.

We define two assumptions about the environment:
\begin{itemize}
\item[] {\em if the agent requests a plan to get to some position,
    then it eventually believes it has received a plan.}
\end{itemize}
\begin{equation}
\begin{array}{c}
\mathsf{PlanningSucceeds}(Pos) \equiv \\
\always (\lactions{ag1}{\mathsf{query}(\mathsf{get\_close\_to}(Pos, P))} \Rightarrow
\sometime \lbelief{ag1}{\mathsf{have\_plan}(Pos, plan)})
\end{array}
\end{equation}

\begin{quote}
{\em if the agent executes a plan
then eventually it believes it has reached the desired position.}
\end{quote}
\begin{equation}
\begin{array}{c}
\mathsf{PlanExecutionSucceeds}(Pos)\quad \equiv \quad
\always (\lactions{ag1}{\mathsf{perf}(\mathsf{execute}(plan))} \Rightarrow
\sometime \lbelief{ag1}{\mathsf{in\_position}(Pos)})
\end{array}
\end{equation}

\noindent We want to establish, using model checking, that:
\begin{itemize}
\item[] {\em if the agent receives a plan on request, and the
    execution of that plan takes it to the middle position in a line,
    then eventually the agent will believe it is maintaining a
    position in the middle of the line.}
\end{itemize}
\begin{equation}
\begin{array}{c}
\mathsf{PlanningSucceeds}(middle) \land
\mathsf{PlanExecutionSucceeds}(middle) \Rightarrow 
\sometime \lbelief{ag1}{\mathsf{maintaining}(middle)}
\end{array}
\label{test1}
\end{equation}

\noindent If we wish to relax our environmental assumptions, we can
also show that
\begin{quote}
{\em if plan execution always succeeds, then either the agent will
eventually believe it is in its desired position, or it never believes
it has a plan for getting there.
}
\end{quote}
\begin{equation}
\begin{array}{c}
\mathsf{PlanExecutionSucceeds}(middle) \Rightarrow \\
\sometime \lbelief{ag1}{\mathsf{maintaining}(middle)} \lor
\always \neg \lbelief{ag1}{\mathsf{have\_plan}(middle, plan)}
\end{array}
\label{test3}
\end{equation}

\begin{table}
\centering
\small 
{\bf Inputs:} 

\begin{tabular}{|l||l|l|l|l|} \hline
Property & close\_to & broken & thruster\_bank\_line & get\_close\_to \\ \hline
\eqref{test1} & middle & $\times$ & $\times$ & (middle, plan) \\
\eqref{test3} & middle & $\times$ & $\times$ & (middle, plan)  \\ \hline
\end{tabular} 
\medskip

{\bf Results:} 

\begin{tabular}{|l||l|l|} \hline
Property & States & Time\\ \hline
\eqref{test1} & 33 & 10s\\
\eqref{test3} & 33 & 10s\\ \hline
\end{tabular} \\
\caption{Results for Analysis of a Single agent with no Thruster Failure}
\label{results:single_nothruster}
\end{table}

\noindent Table~\ref{results:single_nothruster} shows the environment
provided for each property (Inputs), and the size of the generated product automata (in states) and the time taken in hours, minutes and seconds on a dual core 2.8GHz Macbook with 8GB of memory running MacOS X 10.7.5\footnote{It can be seen that time taken does not scale consistently with the number of states in the product automata.  This is because the time taken to generate the automata was sensitive to the complexity of generating new states within the agent code -- for instance, where a lot of Prolog-style reasoning was needed there was a significant slow down in the generation of new program states.} (Results).  The columns on the left show the arguments that 
were supplied to each of the potential percepts (i.e., every time it
took an action the agent could either gain the perception
{\lstinline{get_close_to(middle, plan)}} or not, but could not,
for instance, get the perception
{\lstinline{get_close_to(left, plan)}}).  $\times$ indicates that
the percept was not supplied at all.

These results could be combined deductively with results about the
correct performance of the Continuous and Physical engines to produce
a proof that the the agent would eventually believe it was in
position. Furthermore, with additional analysis to prove that the
sensors always operated correctly:
\begin{quote}
{\em if the agent believes it is in
the middle then it really is in the middle}
\end{quote}
\begin{equation}
\lbelief{ag1}{\mathsf{maintaining}(middle)} \Rightarrow \mathsf{in\_position}(ag1, middle)
\end{equation}
then the theorem could be extended to prove that the agent
really could be guaranteed always to reach its position.

The important thing to note is that the theorem resulting from the
model checking process explicitly states its assumptions about the
real world (i.e., that plans are always produced and are always
accurate) rather than concealing these assumptions within the coding
of the model.  Since it is, in reality, unlikely that you could ever
guarantee that plan execution would always succeed it might be
necessary to combine the model checking with probabilistic results to
obtain an analysis of the likely reliability of the agent (see
further work in Section~\ref{sec:concl}).

\paragraph{Investigating the Response to Thruster Failure}

If we include the possibility of thruster failure into our analysis
then we can show that
\begin{quote}
{\em if the planning process succeeds then either the agent eventually
  believes it is maintaining the position or it
believes it has a broken thruster.}
\end{quote}
\begin{equation}
\begin{array}{c}
\mathsf{PlanningSucceeds}(middle) \land \mathsf{PlanExecutionSucceeds}(middle) \Rightarrow \\
\sometime \lbelief{ag1}{\mathsf{maintaining}(middle)}
\lor \\ \lbelief{ag1}{\mathsf{broken}(x)} \lor 
\lbelief{ag1}{\mathsf{broken}(y)} \lor \lbelief{ag1}{\mathsf{broken}(z)}
\end{array}
\label{test4}
\end{equation}

\noindent We can improve on this result by adding extra assumptions
about the way the environment behaves:
\begin{quote}
{\em whenever the agent fixes a fuel line then eventually it believes the thruster
is working again.}
\end{quote}
\begin{equation}
\begin{array}{c}
\mathsf{ChangingLineSucceeds}(T)\quad \equiv \quad
 \always (\lactions{ag1}{\mathsf{perf}(\mathsf{change\_line}(T))} \Rightarrow
\sometime \neg \lbelief{ag1}{\mathsf{broken}(T)})
\end{array}
\end{equation}

\noindent where $T$ is the thruster effected. A second hypothesis is 
\begin{quote}
{\em broken thrusters never lead to an abort because of thruster failure.}
\end{quote}
\begin{equation}
\begin{array}{c}
\mathsf{NoIrrepairableBreaks}(T)\quad \equiv \quad
\always (\lbelief{ag1}{\mathsf{broken}(T)} \Rightarrow
\always \neg \lbelief{ag1}{\mathsf{aborted(thruster\_failure)}})
\end{array}
\end{equation}

\noindent Property~\eqref{test5} states that 
\begin{quote} {\em if planning succeeds for the middle position, and
    the x thruster is always believed to be fixable and changing a
    fuel line means eventually the agent believes the thruster is no
    longer broken, then eventually the agent will believe it is
    maintaining its position.}
\end{quote}
\begin{equation}
\begin{array}{c}
\mathsf{PlanningSucceeds}(middle) \land
\mathsf{PlanExecutionSucceeds}(middle) \\
 \land\ \mathsf{ChangingLineSuceeds}(x)
\land\ \mathsf{NoIrrepairableBreaks}(x) \\
\Rightarrow
\sometime \lbelief{ag1}{\mathsf{maintaining}(middle)}
\end{array}
\label{test5}
\end{equation}
This is obviously only true if the {\lstinline{y}} and {\lstinline{z}}
thrusters do not also break.  Unfortunately expressing the conditions
for the additional thrusters made the automaton too large for
construction in reasonable time (however, see `aside' below).  As a
result this property was checked only with the option of sending the
agent the predicate $\mathsf{broken(x)}$ not with the option of
$\mathsf{broken(y)}$ or $\mathsf{broken(z)}$.  However this
restriction is clear from the presentation of results in
Table~\ref{results:single}.

\begin{table}
\centering
\small
{\bf Inputs:} 

\begin{tabular}{|l||l|l|l|l|} \hline
Property & close\_to & broken & thruster\_bank\_line & get\_close\_to \\ \hline
\eqref{test4} & middle & x, y, z & (x, 1, 1), (y, 1, 1), (z, 1, 1) & middle \\
\eqref{test5} & middle & x & (x, 1) & middle \\ \hline
\end{tabular} 
\medskip

{\bf Results:} 

\begin{tabular}{|l||l|l|} \hline
Property
& States & Time\\ \hline
\eqref{test4} & 16,609 & 1h,18m,42s \\
\eqref{test5} & 2,890 & 7m,19s \\ \hline
\end{tabular}
\caption{Results of the Analysis of a Single Agent with Thruster Failures}
\label{results:single}
\end{table}


\subsubsection{Analysis of Multi-Agent Models and Communication}

Appendix~\ref{leo:leader} shows the code for a lead agent which determines the position of other agents in a formation.

In the previous examples we were able to determine the appropriate
random inputs to the verification by inspecting the shared beliefs
asserted by the Abstraction Engines for each agent.  The lead agent
has no abstraction engine since it is not controlling a physical
system directly and so does not actually receive any information from
shared beliefs.  It receives all its inputs as messages from other
agents.  We therefore have to analyse it in the light of the messages
the other agents in the system (e.g., {\lstinline{ag1}}) send
rather than by analysis of the Abstraction Engine.  Messages are
passed directly between rational agents rather than being processed by
the Abstraction Engines.
Just as we ``read off'' the shared beliefs from the Abstraction Engine
we can ``read off'' possible messages from the Reasoning Engine code
of the other agents in the system. However, this does not allow us to
verify the system against the presence of unknown agents, or message
corruption.  It is important to note, therefore, that we are
explicitly \emph{not} investigating this in the context of a system
where malicious agents may send conflicting messages.

Using this analysis we determine the following messages that may be
received by agents in the system.  \lstinline{aglead} may receive the
message \lstinline{maintaining(AgName)} in which agent,
\lstinline{AgName}, asserts that is is maintaining its position and
the message \lstinline{aborted(thruster_failure, AgName)} in which
agent, \lstinline{AgName} states that it has aborted its current
maneouvre because of thruster failure.  The satellite agents (as
represented by \lstinline{ag1}) can receive three messages,
\lstinline{assuming_formation(F)} which tells them that the group will
be assuming some formation, \lstinline{F}; \lstinline{position(Pos)}
which informs them that they should assume position, \lstinline{Pos}
within the formation; and the instruction
\lstinline{drop_formation(F)} which tells to to abandon the attempt to
assume the formation \lstinline{F}.

Where, in the system under consideration,
{\lstinline{AgName}} can be {\lstinline{ag1}},
{\lstinline{ag2}}, {\lstinline{ag3}} or
{\lstinline{ag4}}.  \textsf{F} is one of
{\lstinline{line}} or {\lstinline{square}}, and
\textsf{Pos} could be {\lstinline{none}},
{\lstinline{right}}, {\lstinline{middle}},
{\lstinline{left}}, {\lstinline{topright}},
{\lstinline{topleft}}, {\lstinline{bottomright}} or
{\lstinline{bottomleft}}.  For reasons of efficiency, we do not
consider the issue of thruster failure in the multi-agent system so we
will not use values of 
{\lstinline{aborted(thruster_failure, AgName)}} in what follows. 

\paragraph{Single Formation}

We verified the behaviour of both the leader agent and the follower
agents in the situation where there is only one formation, a line, to
be adopted. We needed a new assumption about the performance of the
environment, stating that:
\begin{quote}
\em once the leader believes it has
informed an agent that it should assume a position in a line formation then,
eventually, it will believe it has received a message telling it that the (informed)
agent is maintaining that position.
\end{quote}
\begin{equation}
\begin{array}{c}
\mathsf{AlwaysResponds}(AgName, Pos) \equiv \\
\always(\lbelief{aglead}{\mathsf{informed}(AgName, Pos)} \Rightarrow \sometime
\lbelief{aglead}{\mathsf{maintaining}(AgName))}
\end{array}
\end{equation}

\noindent With this assumption we were able to verify that
\begin{itemize}
\item[] {\em if all agents respond, the lead agent eventually believes
    the agents have assumed a linear formation.}
\end{itemize}
\begin{equation}
\begin{array}{c}
\mathsf{AlwaysResponds(}ag1, line\mathsf{)} \land
\mathsf{AlwaysResponds(}ag2, line\mathsf{)} \land \\
\mathsf{AlwaysResponds(}ag3, line\mathsf{)} \land \mathsf{AlwaysResponds(}ag4, line\mathsf{)} \\
\Rightarrow
\sometime \lbelief{aglead}{\mathsf{in\_formation}(line)}
\end{array}
\label{test6}
\end{equation}

\noindent We also verified some \emph{safety} properties, e.g.: 
\begin{quote}
{\em the leader never believes it has assigned an agent (ag1 in the
  case shown below) to two
  positions at the same time.}
\end{quote}
\begin{equation}
\begin{array}{c}
\always \lbelief{aglead}{\mathsf{position}(ag1, \mathit{left})} \Rightarrow \\ \lnot
(\lbelief{aglead}{\mathsf{position}(ag1, middle)} \land
\lbelief{aglead}{\mathsf{position}(ag1, right)})
\end{array}
\label{test16}
\end{equation}
\begin{quote}
{\em the leader never believes it has assigned two agents to the same
  position (the left in the case shown below).}
\end{quote}
\begin{equation}
\begin{array}{c}
\always \lbelief{aglead}{\mathsf{position}(ag1, \mathit{left})} \Rightarrow \\
\lnot
(\lbelief{aglead}{\mathsf{position}(ag2, \mathit{left})} \lor
\lbelief{aglead}{\mathsf{position}(ag3, \mathit{left})} \lor \\
\lbelief{aglead}{\mathsf{position}(ag4, \mathit{left})})
\end{array}
\label{test17}
\end{equation}

\noindent The follower agent uses the code investigated in our single
agent case, but when it is interacting with a multi-agent system we
want to verify that the messages it sends to the leader agent
accurately portray its beliefs. So, we show that
\begin{quote}
{\em under the assumption that planning and plan execution are
successful for the relevant formation and position, the follower will
eventually believe it has informed the leader that it is maintaining 
its position.}
\end{quote}
\begin{equation}
\begin{array}{c}
\mathsf{PlanningSucceeds}(middle) \land \mathsf{PlanExecutionSucceeds}(middle)  \Rightarrow \\
\always (\lbelief{ag1}{\mathsf{handling}(\mathit{assuming\_formation}(line))} \land
\lbelief{ag1}{\mathsf{my\_position\_is}(middle)}
\Rightarrow \\
\sometime \lbelief{ag1}{\mathsf{sent}(aglead, \mathsf{maintaining}(ag1))}
\end{array}
\label{test7}
\end{equation}

\noindent We can also verify that 
\begin{quote}
{\em followers only send messages if they believes
they are maintaining the positions they have been assigned.}
\end{quote}
\begin{equation}
\begin{array}{c}
\always (\lactions{ag1}{\mathsf{send}(aglead, \mathsf{maintaining}(ag1))} \Rightarrow \\
\lbelief{ag1}{\mathsf{my\_position\_is}(middle)} \land \lbelief{ag1}{\mathsf{maintaining}(middle)})
\end{array}
\label{test11}
\end{equation}

\begin{table}
\small
\centering
{\bf Leader Agent}

\begin{tabular}{|l||l|l||l|l|} \hline
Property & maintaining & aborted & States & Time\\ \hline
$\eqref{test6}$ & ag1, ag2, ag3, ag4 & $\times$ & 1,381 & 18m, 6s\\ \hline
$\eqref{test16}$ & ag1, ag2, ag3, ag4 & $\times$ & 1,563 & 10m, 53s\\ \hline
$\eqref{test17}$ & ag1, ag2, ag3, ag4 & $\times$ & 1,563 & 10m, 53s\\ \hline
\end{tabular}

\medskip

{\bf Follower Agent (Inputs)}

\begin{tabular}{|l||l|l|l|l|l|l|} \hline
Property & close\_to & get\_close\_to &
assume\_formation & position & drop\_formation  \\ \hline
\eqref{test7} & middle & (middle, plan) & line & $\times$
& line 
\\ \hline
\eqref{test11} & middle & (middle, plan) & line & $\times$ 
 & line   \\ \hline
\end{tabular}

\medskip

{\bf Follower Agent (Results)}

\begin{tabular}{|l||l|l|} \hline
Property
& States & Time \\ \hline
\eqref{test7} & 472 & 3m,34s
\\ \hline
\eqref{test11} & 472 & 3m,15s \\ \hline
\end{tabular}

\caption{Analysis of a Multi-Agent system with no thruster failure
attempting to move into a line}
\label{results:multi_nofailure_line}
\end{table}

\noindent The results of this analysis are shown in Table~\ref{results:multi_nofailure_line}

\paragraph{Changing Formations}
Lastly we investigated the behaviour of the leader agent in situations
where the formation could change.  


\begin{quote}
{\em if all agents respond, then
  eventually the leader agent will believe a square formation to have
  been achieved.}
\end{quote}
\begin{equation}
\begin{array}{c}
\mathsf{AlwaysResponds(}ag1, square\mathsf{)} \land
\mathsf{AlwaysResponds(}ag2, square\mathsf{)} \land \\
\mathsf{AlwaysResponds(}ag3, square\mathsf{)} \land \mathsf{AlwaysResponds(}ag4, square\mathsf{)} \\
\Rightarrow
\sometime(\lbelief{aglead}{\mathsf{in\_formation}(square)})
\end{array}
\label{test13}
\end{equation}

\begin{quote}
{\em if all agents respond, then whenever
  the leader agent believes all the agents to be in square formation
  it will 
  eventually believe them to be in a line formation.}
\end{quote}
\begin{equation}
\begin{array}{c}
\mathsf{AlwaysResponds(}ag1, line\mathsf{)} \land
\mathsf{AlwaysResponds(}ag2, line\mathsf{)} \land \\
\mathsf{AlwaysResponds(}ag3, line\mathsf{)} \land \mathsf{AlwaysResponds(}ag4, line\mathsf{)} \\
\Rightarrow 
\always(\lbelief{aglead}{\mathsf{in\_formation}(square)} \Rightarrow \sometime(\lbelief{aglead}{\mathsf{in\_formation}(line)}))
\end{array}
\label{test14}
\end{equation}

\begin{table}
\small
\centering
\begin{tabular}{|l||l||l|l|} \hline
Property & maintaining & States & Time \\ \hline
\eqref{test13} & ag1, ag2, ag3, ag4  & 1,892 & 29m,58s\\ \hline
\eqref{test14} & ag1, ag2, ag3, ag4  & 3,333 & 1h,5m,21s\\ \hline
\end{tabular}
\caption{Analysis of results of a multi-agent system with no failures
but changing formations}
\label{results:multi_change}
\end{table}

\noindent The results of this are shown in
Table~\ref{results:multi_change}
\bigskip

\noindent We can, of course, continue with further
verification. However, it should be clear to the reader by now how
this proceeds, combining model checking of behaviour for individual
agents in the presence of a random environment, together with
straight-forward temporal/modal reasoning that can be carried out by hand
or by appropriate automated proof tools. 





\section{Scenario:~Adaptive Cruise Control}
\label{sec:adaptive}

For our final example we demonstrate how our approach can integrate
with approaches based on hybrid automata or hybrid programs which
focus on the continuous dynamics of the system.  We look at an example
developed in \keymaera\ by~\cite{sarah11:_adapt}.  This example
considers the problem of a car with adaptive cruise control that must
maintain a safe distance between itself and the car in front and may
only change lane when it is safe to do so.  Loos et al., analyse this
problem compositionally working up from the simple case of two cars in
a single lane to an arbitrary number of cars on a motorway with an
arbitrary number of lanes.  The control system for the cars is
modelled as a hybrid program.

It is outside the scope of this paper to describe the syntax and
semantics of hybrid programs.  But we reproduce a simple example
from~\cite{sarah11:_adapt} to give a flavour of the language.
\begin{figure}[htbp]
\begin{eqnarray}
\mathsf{llc} & \equiv & (ctrl;dyn)* \nonumber \\
ctrl & \equiv & l_{ctrl} || f_{ctrl} \nonumber \\
l_{ctrl} & \equiv & (a_l ::= *; ?(-B \leq a_l \leq A)) \nonumber \\
f_{ctrl} & \equiv & (a_f ::= *; ?(-B \leq  a_f \leq -b)) 
~~\cup~~ (?\mathbf{Safe}_{\epsilon}; a_f ::= *; ?(-B \leq a_f \leq A)) 
~~\cup~~ (?(v_f = 0); a_f ::= 0) \nonumber \\
 dyn & \equiv & (t:=0; x'_f = v_f, v'_f = a_f, x'_l = v_l, v'_l = a_l, t' = 1, v_f \geq 0 \wedge v_l \geq 0 \wedge t \leq \epsilon) \nonumber
\end{eqnarray}
\caption{Hybrid Program for a Leader and a Follower Car in a Single
  Lane.}
\label{fig:lfcar}
\end{figure}
In Fig.~\ref{fig:lfcar}, $x_f$ is the position of the follower car,
$v_f$ its velocity and $a_f$ its acceleration.  Similarly $x_l, v_l$
and $a_l$ represent the position, velocity and acceleration of the
leader car.  $-B$ is a global constant for the maximum possible
deceleration due to braking, $-b$ is a global constant for the minimum
deceleration and $A$ is the maximum possible acceleration.  $\epsilon$
is an upper bound on the reaction time for all vehicles.
$\mathbf{Safe}_{\epsilon}$ is an equation used by the control system
to determine the safe distance between cars.  A key part of the
verification effort reported is establishing this equation and proving
that it does guarantee the system to be collision free.  It is defined
as

\begin{equation}
\mathbf{Safe}_{\epsilon} \equiv x_f + \frac{v^2_f}{2b} + \left( \frac{A}{b} + 1 \right)\left(\frac{A}{2}\epsilon^2 \right) < x_l + \frac{v_l^2}{2B}
\end{equation}

\noindent In Fig.~\ref{fig:lfcar}, $\mathsf{llc}$ defines a repeated
sequence of a discrete control, $ctrl$, followed by dynamic control,
$dyn$.  The discrete control is the parallel composition of the
control of the leader, $l_{ctrl}$, and the follower, $f_{ctrl}$.  The
leader has a simple control program --- it may adopt any acceleration
($a_l ::= *$) within the available range.  $f_{ctrl}$ defines three
possible states.  In the first $(a_f ::= *; ?(-B \leq a_f \leq -b))$
the acceleration may be any negative value.  In the second $
(?\mathbf{Safe}_{\epsilon}; a_f ::= *; ?(-B \leq a_f \leq A))$ the
acceleration may be any possible value provided the car is a safe
distance behind the leader, and in the final state $(?(v_f = 0); a_f
::= 0)$, the acceleration may be zero if the velocity is zero.  The
dynamic control, $dyn$, states how position and velocity vary
depending upon acceleration, using the standard differential equations
for motion.

The control systems for the cars in the more complex examples follow a
similar form -- represented as a non-deterministic choice over legal
states for the system.  The decision-making represented by the
constraints on each state is relatively simple here with no reference,
for instance, to any goals of the system.  It would be possible to
represent more complex control in hybrid programs, particularly since
the constraints have access to full first order logic (unlike the
constraints in linear hybrid automata) as well as simple if-then
control structures, but it would be cumbersome to do so since the
language remains low level.

We implemented an agent for adaptive cruise control within the system
described in Section~\ref{sec:eass:arch}.  This adopted the above
rules for safety but added in additional features involving goal-based
and normative behaviour.  We present the code for the Reasoning Engine
here, the code for the Abstraction Engine can be found in
Appendix~\ref{adaptive:abs}.

The simple case of a car travelling in a single lane of a motorway is
shown in Listing~\ref{code:glc}.  The agent has a goal to drive at the
speed limit.  To achieve this goal it accelerates if it believes it
can do so and then waits for a period before checking if the goal is
achieved.  If it can not accelerate then it waits until it believes it
can. The construct $*b$ causes an intention to suspend until $b$
becomes true.  \lstinline{can_accelerate} is determined by a belief
rule and is true if it is safe to accelerate and the driver is not
currently taking any action.  Once the car has reached the speed limit
(line 18) it maintains its speed.  At this point the goal will be
dropped because it has been achieved.  So if the speed drops below the
speed limit the goal will be re-established (lines 20, 21).  If it
stops being safe to accelerate, the agent brakes (line 23).  Actions
by the driver override decisions by the agent, but it still will not
accelerate unless it believes it is safe to do so.


We tested the agent in a simple \java\ environment which modelled the
dynamics described by~\cite{sarah11:_adapt} and performed the
calculation of $\mathbf{Safe}_{\epsilon}$ passing the perception,
\lstinline{safe}, to the Abstraction Engine if
$\mathbf{Safe}_{\epsilon}$ returned true\footnote{We could have had
  the Abstraction Engine calculate `safe' but it seemed more in
  keeping with~\cite{sarah11:_adapt} to have this calculated
  centrally.}.

We then used our agent based model checking to verify that the agent
obeyed the constraints verified by Loos et al. (i.e., it only
accelerated if it was safe to do so) We analysed the abstraction
engine (shown in Appendix~\ref{adaptive:abs}) to determine the
possible incoming shared beliefs were \lstinline{safe},
\lstinline{at_speed_limit}, \lstinline{driver_accelerates},
and \lstinline{driver_brakes}.  Then we proved that
\begin{quote}
  \emph{whenever the agent accelerates then it believes itself to be a
    safe distance from the car in front:}
\end{quote}
\begin{equation}
  \always (\lactions{\mathit{car}}{\mathsf{accelerate}} 
  \rightarrow 
  \lbelief{\mathit{car}}{\mathsf{safe}})
\end{equation}

\begin{lstlisting}[float,caption=Cruise Control Agent (Single Lane):Reasoning Engine,basicstyle=\sffamily,style=easslisting,language=Gwendolen,label=code:glc]
:Rules:

B can_accelerate :- B safe & lnot B driver_accelerates & lnot B driver_brakes;

:Initial Goals:

at_speed_limit [achieve]

:Plans:

+! at_speed_limit [achieve] : {B can_accelerate} <-
       perf(accelerate),
       wait;
+! at_speed_limit [achieve] : {lnotB can_accelerate} <- *can_accelerate;

+at_speed_limit: {B can_accelerate, B at_speed_limit} <- 
        perf(maintain_speed);
-at_speed_limit: {lnotG at_speed_limit [achieve], lnotB at_speed_limit} <-
        +! at_speed_limit[achieve];

-safe: {lnotB driver_brakes, lnotB safe} <- perf(brake);

+driver_accelerates: {B safe, lnotB driver_brakes, B driver_accelerates} <- 
       perf(accelerate);
+driver_brakes: {B driver_brakes} <- perf(brake);
\end{lstlisting}

\noindent We also investigated a more complex case (also analysed by
Loos et al) in which cars could change lane.  We implemented the
normative rules of the UK highway code so that the agent would always
attempt to drive in the leftmost lane unless it wished to overtake a
car in front, in which case it would attempt to overtake on the right.

The code is shown in Listing~\ref{code:ghc}.  This code introduces
further features of our \gwendolen\ language variant.  The action
`{\lstinline{perf}}' is interpreted by the environment as a message to
the Abstraction Engine to adopt a goal which in turn cause the
execution of some non-symbolic code.  $+_{\Sigma}${\lstinline{b}} and
$-_{\Sigma}${\lstinline{b}} are used to indicate when an agent adds
{\lstinline{b}} to, or removes {\lstinline{b}} from, the set of shared
beliefs.  We introduce a new type of \emph{perform} goal which is
distinguished from an achieve goal by the fact that there is no check
that it succeeds; once the actions in a plan associated with a perform
goal have been executed then the goal vanishes.  Lastly,
\lstinline{.lock} and \lstinline{.unlock} are used to \emph{lock} an
intention to remain current, this ensures that the sequence of deeds
are executed one after the other without interference from other
intentions.

The code represents an agent which can be in one of two contexts,
moving left or overtaking which are represented as beliefs.  Lines
9-10 control switching between these contexts.  The use of contexts
allows us to control dropping goals as the situation changes even if
they have not been achieved.  The percepts \lstinline{car_ahead} and
\lstinline{car_ahead_in_left_lane} are supplied by the Abstraction
Engine and represent the detection of a car the agent wishes to
overtake or pass on the right.  The intention is that these beliefs
are shared \emph{before} the car would need decelerate in order to
maintain a safe distance from the car in front (although we verified
the more general case where the beliefs could appear in any order).
Similarly \lstinline{clear_left} is used by the agent to indicate
whether there is no car in the left hand lane which the agent wants to
overtake before moving left.  Lines 13-15 control the assertion and
removal of \lstinline{clear_left}.  Lines 16-17 control the adoption
and abandonment of the goal, \lstinline{in_leftmost_lane}.  Lines
19-26 are the code for achieving this goal depending upon whether or
not it is safe to move left, or there is a car in the left hand lane
to be overtaken before moving left.  Lines 28-45 work similarly for
the \lstinline{overtaking} context with the belief
\lstinline{overtaken} used by the agent to keep track of whether it
has successfully overtaken a car.

\begin{lstlisting}[float,caption=Cruise Control Agent (Changing Lanes):Reasoning Engine,basicstyle=\sffamily,style=easslisting,language=Gwendolen,label=code:ghc]
:Initial Goals:

in_leftmost_lane [achieve]

:Plans:

// Switching Context
+! switch_overtake [perform] : {True} <- +.lock,-moving_left,+overtaking,-.lock;
+! switch_move_left [perform] : {True} <- +.lock,-overtaking,+moving_left,-.lock;

// Moving left
-car_ahead_in_left_lane: {~B clear_left, lnotB car_ahead_in_left_lane} <- 
    +clear_left;
+car_ahead_in_left_lane: {B clear_left, B car_ahead_in_left_lane} <- -clear_left;
+moving_left: {B moving_left} <- +! in_leftmost_lane[achieve];
-moving_left: {G in_leftmost_lane[achieve]} <- -!in_leftmost_lane[achieve];

+!in_leftmost_lane [achieve]: 
    {B safe_left, lnotB car_ahead_in_left_lane, B moving_left} <- 
        +.lock, remove_shared(changed_lane), perf(change_left), -.lock, *changed_lane;
+! in_leftmost_lane [achieve]: {lnotB safe_left, B moving_left} <- *safe_left;
+! in_leftmost_lane [achieve]: {B car_ahead_in_left_lane, B moving_left} <- 
    *clear_left;
+! in_leftmost_lane [achieve]: {lnot B moving_left} <- 
    -!in_leftmost_lane[achieve];

// Overtaking
+car_ahead_in_lane: {B moving_left, B car_ahead_in_lane, lnotB in_rightmost_lane} 
    <- -overtaken, +! switch_overtake[perform];
+overtaking: {B overtaking} <- +! overtaken [achieve];
-overtaking: {G overtaken[achieve]} <- -!overtaken[achieve];

+! overtaken [achieve]: 
    {B safe_right, B car_ahead_in_lane, B overtaking, lnotB in_rightmost_lane} <-
	+.lock, remove_shared(changed_lane), perf(change_right), -.lock, 
        *changed_lane, +overtaken;
+! overtaken [achieve]: 
    {lnot B safe_right, B car_ahead_in_lane, B overtaking, lnotB in_rightmost_lane} 
        <- *safe_right;
+! overtaken [achieve]: {lnot B car_ahead_in_lane, B overtaking} 
    <- +! switch_move_left[perform];
+! overtaken [achieve]: {lnotB overtaking} <- -! overtaken[achieve];
+! overtaken [achieve]: {B in_rightmost_lane, B overtaking} 
    <- +! switch_move_left[perform];
\end{lstlisting}

Once again we analysed the abstraction engine (Appendix~\ref{adaptive:abs}) to obtain the list of shared beliefs that might be sent to the rational engine: \lstinline{in_leftmost_lane}, \lstinline{in_rightmost_lane}, \lstinline{changed_lane}, \lstinline{safe_right}, \lstinline{safe_left}, \lstinline{car_ahead_in_lane}, \lstinline{car_ahead_in_left_lane}.

In this system we were able to verify that 

\begin{quote}
  \emph{the car only changes lane to the right if it believes it is
    safe to do so and the car only changes lane to the left if it
    believes it is safe to do so:}
\end{quote}
\begin{equation}
\always (\lactions{\mathit{car}}{\mathsf{change\_right}} \rightarrow \lbelief{\mathit{car}}{\mathsf{safe\_right}}) \wedge
\always (\lactions{\mathit{car}}{\mathsf{change\_left}} \rightarrow \lbelief{\mathit{car}}{\mathsf{safe\_left}}) 
\end{equation}

\noindent It should be noted that our verification here shows that the
implemented rational agent adheres to the constraints on the discrete
control used in the verification of the full hybrid system by Loos et
al.  We assume, among other things, that `\lstinline{safe}' is
calculated correctly.  The point is not to redo the safety analysis
but to show that the additional normative and goal-directed behaviours
do not effect the agent's ability to obey the safety constraints.

\section{Concluding Remarks}
\label{sec:concl}
In this paper we have presented a compositional approach to the
verification of the decision-making parts of hybrid autonomous
systems, demonstrating both that such an approach is feasible and
how it can be pursued.
The methodology uses model checking to 
perform formal verification of the reasoning of a rational agent
that is controlling a hybrid autonomous system.  This approach avoids
the necessity for complex and opaque models of the external system.
At the same time our approach allows the results of model checking to be
combined deductively (in a straight-forward way) with results about
the full system, in order to provide formal analysis of the whole.

Our hypothesis is that different verification techniques are more
suited to different aspects of hybrid system analysis.  Program model
checking is well-adapted to the analysis of rational agent code, but
less well adapted to reasoning in continuous domains where
probabilistic, analytical and deductive techniques may yield more
accurate results.  Our approach allows theorems about the rational
agent to be combined deductively in a compositional fashion with
results about the rest of the system.  Our claim is that it is easier,
under this methodology, to make explicit the assumptions we have
utilized in order to allow verification to take place.  While we have
focused particularly on the program model checking of rational agent
implementations, the methodology would also be suitable for general
model checking of complex agent-based (and possibly other) {\em
  models} for discrete control of hybrid systems.  This could be done
in a specialised model checker for agent systems such as
\mcmas~\cite{lomuscio09:_mcmas} or \mck~\cite{gammie04:_mck}.  In
essence we believe approaches such as ours are suitable wherever there
is significant complexity within the discrete reasoning required to
control a hybrid system.

To exhibit this approach, we explored three examples in which
verification was performed in this compositional manner.  The first
example used a simple program based on the RoboCup Rescue scenario.
Our second example examined code actually developed separately
for the control of satellites in low Earth orbit, and our third looked at an adaptive cruise control system already analysed in \keymaera.  In all examples
we were able to prove properties about the efficacy of the system in
the absence of a detailed formal analysis of real world (and
continuous) aspects. Work of this kind enables us to provide accurate
and clear analysis of the autonomous decisions made by agents in real
world scenarios.  It is made feasible by the ability for us to refer to the beliefs, goals and actions of the agent in the property specification language, which allow us to limit our theorems to the internals of the agent.

Our interest is primarily in the verification of discrete logic/agent
based decision-making algorithms for controlling hybrid systems,
rather than in the analysis of the underlying continuous control.  As
such our approach abstracts away from the continuous aspects entirely,
much as many hybrid automata based analyses abstract away from the
details of decision-making.  This potentially leads to
over-approximation, where we examine all the inputs that an agent
could receive not just those that ``make sense'' from the continuous
perspective -- e.g., in our adaptive cruise control example we
explored the case where the driver is reported to be both braking and
accelerating at the same time.  It is a matter of some debate whether
this truly represents an over-approximation since it would potentially
be possible for a malfunctioning sensor to transmit contradictory
information.  However it is clear that the over-approximation also has
repercussions on the search space as the number of potential sensor
inputs increases which, combined with the inefficiencies of program
model checking currently limits the scope of the methodology.

There are a number of avenues for further work. The most urgent
problem is that the agent verification technology is actually very
\emph{slow} (as can be seen from the times reported in
Section~\ref{sec:case_study}). Program model checking, such as that
provided by Java Pathfinder (\jpf), analyzes the real code, often in
an explicit way which increases the time taken to generate new states
in the model.  Improving the efficiency of \ajpf, including the
investigation of well-known abstraction techniques such as
property-based slicing~\cite{BFWV09:Slicing}, is the subject of
on-going work.  Following the work
by~\cite{hunter13:_syner_exten_framew_multi_agent_system_verif}, we
have also investigated the use of \ajpf\ to generate models of the
program that can be exported into other model checking systems such as
\spin, NuSMV or \mcmas~\cite{dennis13:_using_agent_jpf_to_build}.

We have shown how agent decision-making can be formally verified in
the presence of random environmental interactions. However, we would
also like to incorporate a more refined \emph{probabilistic} analysis
of environmental interactions. For instance, it is highly probable
that no guarantee can be supplied that data incoming from sensors is
correct, yet it is highly likely that the sensor will come with a
probabilistic guarantee of reliability.  It is therefore desirable to
extend the kind of analysis we have done here to theorems of the form
``Assuming data from sensors is correct 98\% of the time, and actions
succeed 98\% of the time, then 97\% of the time the agent never
believes it has reached a bad state''.  Initial investigation of a
link between \ajpf\ and the probabilistic \prism\ model
checker~\cite{kwiatkowska02prism} reported
in~\cite{dennis13:_using_agent_jpf_to_build} will allow us to explore
this direction further.

We are
also interested in formalising the composition of the environment used
for this kind of analysis.  We presented an informal overview of how
incoming percepts could be chosen and in the examples we performed
this selection was carried out by hand.  However, we believe that much
could be done to automate the process and so ensure that no potential
inputs are overlooked.  

As noted above, the non-deterministic modelling of sensor input places limits on the ability of our system to scale as the number of sensors, and more importantly sensor values, increase.  It would therefore be beneficial to move in the direction of~\cite{frehse04:_assum_i_o} and represent the continuous parts of the system as an abstract automata, or a similar formulation, which allowed us to control the number of possible states.
This would draw on work such as~\cite{Lomuscio:2010:ARL:1939864.1939883,Pasareanu:2008:LDC:1375435.1375437,silva11:_formal_envir_model_for_multi_agent_system}.

Finally, we also wish to consider agent control systems that can
co-operate together, exchange information (and, potentially, plans)
and, in some cases, learn new plans for solving problems.  Techniques
such as those presented above, if embedded within an agent's own reasoning capabilities, would allow individual agents to assess
whether any new plan violated safety constraints imposed upon it, and
would allow groups of agents to reason about whether their combined
capabilities could ever achieve some goal.  

Similarly we are interested in techniques for providing guarantees that an agent always operates within certain bounds no matter what new plans it is given or learns.  Some initial steps towards this have been investigated in~\cite{dennis13:_ethic_choic_in_unfor_circum}.

\begin{small}
\paragraph{Acknowledgements}This research was partially funded by
Engineering and Physical Sciences Research Council grants EP/F037201
and EP/F037570, And primarily occurred while the third and fifth authors
were at University of Southampton.
\end{small}

\bibliographystyle{plain}

\bibliography{eass_verif}

\newpage
\appendix

\section{Code for LEO Satellite Example}
\label{leo}

\subsection{A Satellite Agent}
\label{leo:satellite}
The code in this section is for an agent that is in direct control of a single satellite but which receives information on formations from a separate, coordinating, leader agent.

\subsubsection{The Abstraction Engine}

It should be noted that,
  while we believe a BDI-style programming language to be an
  appropriate implementation language for $R$, the Reasoning Engine, we are investigating the
use of stream processing languages for $A$, the Abstraction Engine.  This is the subject of
ongoing work.  In the example here, however, the abstraction engine is
programmed in our variant of the \gwendolen\ language.

Code for the Abstraction Engine, $A$, is shown in Listing~\ref{code:leoA}.  We use
standard BDI syntax, as described previously, extended with new
constructs for manipulating the shared beliefs. 
$+_{\Sigma}${\lstinline{b}} and
$-_{\Sigma}${\lstinline{b}} are used to indicate when an agent
adds {\lstinline{b}} to
or removes {\lstinline{b}} from the set of shared beliefs.

The code is split into roughly three sections for \emph{abstraction},
\emph{reification} and \emph{housekeeping} (not shown). The key aim
for this satellite is to take up a position in a formation of
satellites. Lines 1--14 provide code for generating abstractions.  As
information about the satellite's current location arrives from $\Pi$
(via `{\lstinline{stateinfo}}') the agent (if it does not
already believe itself to be in location
`{\lstinline{close_to(Pos)}}') requests the calculation of the
distance from its desired location (via
{\lstinline{comp_distance}}) which returns a judgment of
whether the agent is ``close enough'' to the target position. This is
then used to assert a shared belief about whether the agent is in
position or not.

\begin{sloppypar}
Similarly, information about the thruster status is used in lines
8--14.  Thruster information arrives via perception as
`{\lstinline{thruster(X,L1,L2,P,C,V)}}' which gives information
about the pressure on the two incoming fuel lines
({\lstinline{L1}} and {\lstinline{L2}}) and the output
fuel line ({\lstinline{P}}) as well as the current
({\lstinline{C}}) and voltage ({\lstinline{V}}) in
thruster \lstinline{X}.  Based on this information, particularly the observation
that the output fuel pressure has dropped below 1, the agent can assert or
retract shared beliefs about whether a particular thruster is broken
(lines 10 and 12)\footnote{In our implementation interaction with shared
beliefs is handled as an action by the underlying system, changes to
the shared beliefs are then acquired by the perception mechanisms in
the abstraction and reasoning engine.}.
\end{sloppypar}

\begin{sloppypar}
The \emph{reification} code in lines 18--42 tells the abstraction
engine how to respond to requests for calculations or actions from the
Reasoning Engine, $R$, e.g., a request to
`{\lstinline{get_close_to}}' is translated into a calculation
request to $\Omega$ to call the function
`{\lstinline{plan_named_approach_to_location_multi}}'.  The
details of this code are of less interest in terms of verification of
the Reasoning Engine, since they describe its effects upon the rest of
the system rather than the external system's effect on the Reasoning
Engine.  However, it is worth noting that some of these effects still
involve the assertion of new shared beliefs (e.g., in lines 19-22, when the
abstraction engine has a response to its calculation from the
Continuous Engine it asserts the value of this response as a shared
belief {\lstinline{get_close_to(Pos, P)}}).  Similarly it can be
seen that a request to change fuel line
(i.e., `{\lstinline{change_line'}} at line 36--43) involves setting
several valves (`{\lstinline{run}}' takes two arguments, the
name of a \matlab{} function and a list of arguments to be supplied to
that function.  `{\lstinline{pred}}' allows a \matlab{} function
call name to be composed from a number of strings),
 modifying the shared beliefs about which fuel line is
being used (`{\lstinline{thruster_bank_line}}'), and then
waiting for any change to take effect before 
asserting a shared belief that the thruster should no longer be
broken. Note that, at this point, if the pressure is still low on the
output fuel line then the shared belief that the thruster is broken
will be reasserted the next
time thruster data is perceived.
\end{sloppypar}

Finally, the \emph{housekeeping} code handles some details of
abstracting specific thruster information sent by $\Pi$ into more
generic information which passes the name of the thruster as a
parameter, and the code for handling waiting.  This is omitted from
the code fragment shown.

\pagebreak[3]

\begin{lstlisting}[float,caption=Low Earth Orbit: Abstraction Engine,basicstyle=\sffamily,style=easslisting,language=Gwendolen,label=code:leoA]
// Abstraction Code
+stateinfo(X,Y,Z,Xd,Yd,Zd) : { B heading_for(Pos), B position(Pos, Xc, Yc, Zc), 
                               lnotB close_to(Pos) } <-
       comp_distance(X,Y,Z,Xc,Yc,Zc,V),
        +bound(V);
+bound(yes) : { B heading_for(Pos), lnotB close_to(Pos)} <- assert_shared(close_to(Pos));
+bound(no)  : {  B heading_for(Pos), B close_to(Pos)}  <- remove_shared(close_to(Pos));

+thruster(X,L1,L2,P,C,V):
       { B thruster_bank_line(X,N,L), lnotB broken(X), P < 1 } <- assert_shared(broken(X));
+thruster(X,L1,L2,P,C,V):
       { B thruster_bank_line(X,N,L),  B broken(X), 1 < P } <- remove_shared(broken(X));

-broken(X) :
       { B thruster_bank_line(X,N,L), B thruster(X,L1,L2,P,C,V), P < 1 } <-
       assert_shared(broken(X));

// Reification Code
+!get_close_to(Pos, P) : 
       { B position(Pos,Xc,Yc,Zc), B stateinfo(X,Y,Z,Xd,Yd,Zd) } <-
       .calculate(plan_named_approach_to_location_multi(Xc,Yc,Zc,X,Y,Z),P),
       assert_shared(get_close_to(Pos, P));

+!execute(P)    : { B get_close_to(Pos, P) } <-
        +heading_for(Pos), run(pred(set_control), args(P));
+!null  : { True } <- 
        -heading_for(Pos), run(pred(set_control), args("NullOutput"));
+!maintain_path  : { B close_to(Pos) } <-
        -heading_for(Pos),
        run(pred(set_control), args("NullOutput")),
        run(pred(set_control), args("Maintain"));
+!change_thruster(X,N,NewN)  : { B thruster(X,N) } <-
       -thruster(X,N),
       +thruster(X,NewN);

+!change_line(T)  :
        { B thruster_bank_line(T,B,1), B thruster(X,L1,L2,P,C,V,P) } <-
        run(pred("set_",T,"_valves"), args(off,off,on,on)),
        remove_shared(thruster_bank_line(T,B,1)),
        assert_shared(thruster_bank_line(T,B,2)),
        -thruster(X,L1,L2,P,C,V),
        +!wait ,
         remove_shared(broken(T));
\end{lstlisting}
\bigskip

Note that we can read off the changes in the shared beliefs
of the system that may arise simply by looking for the occurrences of
$+_{\Sigma}$ and $-_{\Sigma}$ that occur in the above code.

\subsubsection{The Reasoning Engine}
Code for the Reasoning Engine, $R$, is shown in Code
fragment~\ref{code:leoR}.  We use the same syntax as we did for the
Abstraction Engine, $A$, and for the rescue robots in
Section~\ref{sec:scen2}.  We additionally use the construction
`{\lstinline{*b}}' to indicate that processing of a plan should
suspend until some belief, {\lstinline{b}}, becomes true. Here
the actions, `{\lstinline{perf}}' and
`{\lstinline{query}}', are interpreted by the environment as
messages to $A$ to adopt a goal.  Calls to {\lstinline{query}}
pause execution of the plan in order to wait for a shared belief
indicating a response from the calculation engine.  The plans that are
triggered in the abstraction engine by {\lstinline{perf}} and
{\lstinline{query}} can be seen in the reification part of $A$'s code.

In the Reasoning Engine we use both {\em achieve} and {\em perform}
goals~\cite{RiemsdijkDM09}.  Achieve goals persist in the agent's goal
base until it believes them to be true, so plans for achieve goals may
be enacted several times until they succeed.  Perform goals, on the
other hand, disappear once a plan associated with them has been
executed and contain no implicit check that they have succeeded.

The code for $R$ has several plans and one {\em belief rule}.  A
belief rule is a Prolog-style rule used for deducing beliefs.  In this
case it deduces that a thruster failure is repairable if the thruster
is currently using the first fuel line.

{\lstinline{assuming_formation}} (lines 6--10) is a perform goal that
runs the agent through some general initialisation, i.e., finding the
formation to be assumed (if not already known) and the agent's
position in that formation and then starting an attempt to achieve
that position.  The agent can either start out with the goal of
assuming a formation, or can be sent the goal as a request from
another agent.

Lines 12--20 of code handle instructions from a leader agent to drop the attempt to assume the current formation.

Lines 22--24 deal with converting any new beliefs (typically received
from messages sent by a coordinating leader agent) about the position
the agent is required to adopt ({\lstinline{position}})
into a belief that the agent actually wants to adopt that
position ({\lstinline{my_position_is}}).  Once a position has
been adopted by the agent then no other suggestion is accepted.

{\lstinline{none}} informs the agent that it is not needed for
this formation and so it immediately asserts a belief that it is
maintaining the desired position.

Lines 26--36 handle moving to a position, one case involves the agent
requesting a plan to get there and the other case assumes the agent
already has a plan.

Lines 39--56 handle the agent's plans for getting into a state where it
is maintaining a position, while handling possible aborts.  Once it
believes it is in the right place in the formation it will instruct
the Abstraction Engine to initiate position maintenance procedures,
{\lstinline{perf(maintain_path)}}.  If it is not in the right
place it sets up a goal to get there
{\lstinline{+!in_position(Pos) [achieve]}}.  If a thruster is
broken, but the system has not yet aborted, then the agent waits for
the thruster to be fixed (because {\lstinline{maintaining(Pos)}}
is an achieve goal, once this plan is completed when the thruster is
fixed, the agent will then select a
different plan to achieve the move into position).  Lastly if the system is
aborted the agent will wait for new instructions.  If the satellite
successfully reaches a position then it will perform
{\lstinline{+! cleanup}} to remove any interim beliefs it has
generated such as which plans it has and whether it is attempting to
assume a formation.

Lines 57--60 involve informing a leader agent once the agent is
maintaining its position.  Lines 61--71 handle thruster failures either
by attempting a repair or
by aborting.  Lines 73--85 handle aborts by dropping any goals to
maintain a position, informing a leader agent of the abort and getting
the satellite to cease any attempt to maintain a position.
`{\lstinline{perf(null)}}' switches off the satellites
thrusters, effectively stopping any maneuver it is attempting.

\begin{lstlisting}[float,caption=Low Earth Orbit: Reasoning Engine, style=easslisting, basicstyle=\tiny\sffamily,language=Gwendolen,label=code:leoR]
:Belief Rules:

B repairable(X, with(change_line(X))) :- B thruster_line(X, 1);

:Plans:
+!assuming_formation(F) [perform] : {lnot B assuming_formation(F)} <-
  +!initialise(F) [perform],
  +!my_position_is(X) [achieve],
  +!maintaining(X) [achieve];
+!assuming_formation(F) [perform] : {B assuming_formation(F)};

// May get told to abandon the current formation
+! drop_formation(F) [perform] : {B assuming_formation(F)} <- 
   -! assuming_formation(F) [perform], 
   +! clear_position [perform], 
   +! cleanup [perform],
   perf(null);
+! drop_formation(F) [perform] : {lnot B assuming_formation(F)} <-
   -! assuming_formation(F) [perform],
   perf(null); 

+position(X) : {lnot B my_position_is(Y)} <-   +my_position_is(X);

+my_position_is(none) : {lnot B maintaining(none)} <-  +maintaining(none);

+! in_position(Pos) [achieve] : 
  {lnot B in_position(Pos), lnot B have_plan(Pos, Plan)} <-
  .query(get_close_to(Pos, P)),
  +have_plan(Pos, P),
  perf(execute(P)),
  *close_to(Pos),
  +in_position(Pos);
+! in_position(Pos) [achieve] : {lnot B in_position(Pos), B have_plan(Pos, P)} <-
  perf(execute(P)),
  *close_to(Pos),
  +in_position(Pos);

+! maintaining(Pos) [achieve] : {B in_position(Pos), B assuming_formation(F), lnotB aborted(Reason), lnotB broken(X)} <-
  perf(maintain_path),
  +maintaining(Pos),
  +!cleanup [perform];
+! maintaining(Pos) [achieve] : {lnot B in_position(Pos), B assuming_formation(F),
                                  lnotB aborted(Reason), lnotB broken(X)} <-
  +!in_position(Pos) [achieve],
  perf(maintain_path),
  +maintaining(Pos),
  +!cleanup [perform];
+! maintaining(Pos) [achieve] : {B broken(X), lnotB aborted(Reason)} <- * fixed(X), -fixed(X);
+! maintaining(Pos) [achieve] : {lnotB assuming_formation(F)} <-  -! maintaining(Pos) [achieve];
+! maintaining(Pos) [achieve] : {B aborted(Reason)} <-  * new_instructions(Ins);

+maintaining(Pos) : {B leader(Leader), B my_name(Name)} <-
  .send(Leader, :tell, maintaining(Name)),
  +sent(Leader, maintaining(Name));

+broken(X): {B aborted(thruster_failure)} <- -fixed(X);
+broken(X): {B repairable(X, with(Y)), lnotB aborted(thruster_failure), lnotB fixed(X)} <-
    perf(Y);
+broken(X): {lnotB repairable(X, Y), lnotB aborted(thruster_failure)} <-
   -fixed(X),
   +! abort(thruster_failure) [perform];
+broken(X): {B repairable(X, Y), B fixed(X), lnotB aborted(thruster_failure)} <-
   -fixed(X),
   +! abort(thruster_failure) [perform];
-broken(X): {True} <- +fixed(X);

+!abort(R) [perform]: {B leader(Leader), B my_name(Name), G maintaining(Pos) [achieve]} <-
  +aborted(R),
  -! maintaining(Pos) [achieve],
  .send(Leader, :tell, aborted(R, Name)),
  +send(aborted(R, Name), Leader),
  perf(null);
+!abort(R) [perform]: {B leader(Leader), B my_name(Name),  lnotG maintaining(Pos) [achieve]} <-
  +aborted(R),
  .send(Leader, :tell, aborted(R, Name)),
  +send(aborted(R, Name), Leader);
\end{lstlisting}

We have omitted from the code the initial beliefs and goals of the
agent.  The configuration of those beliefs and goals
creates a number of different agents which we used during testing.  For
instance, if
the agent already has beliefs about the formation that has been chosen
and its position within that formation then it does not need to
request information from the leader agent.
Some initialisation and clean up code has also been omitted.

\medskip

\noindent The architecture lets us represent the high-level decision
making aspects of the system in terms of the beliefs and goals of the
agent and the events it observes. So, for instance, when the
Abstraction Engine, $A$, observes that the thruster line pressure has
dropped below 1, it asserts a shared belief that the thruster is
broken.  When the Reasoning Engine, $R$, observes that the thruster is
broken, it changes fuel line.  This is
communicated to $A$, which then sets the appropriate valves and
switches in the Physical Engine, $\Pi$.

\subsection{A Leader Agent}
\label{leo:leader}

Listing~\ref{code:leoMAS} shows a lead agent,
{\lstinline{aglead}}, which determines the position of other
agents in a formation. This agent has several belief rules
that are designed to establish that all agents have been informed of
their positions in the formation
({\lstinline{all_positions_assigned}} is true if there is no
position in the formation which is not the position of an agent).  The
belief rule for {\lstinline{desired_formation}} has two
configurations: one in which the only formation to be attempted is a
line; and the other in which the agent will start out attempting a
square, and then change to a line.  The belief
{\lstinline{one_formation}} determines which configuration is
adapted and can be used to generate different agents from the same
code.

Lines (15-24) handle the selection of a formation to be adopted and
any clean-up of old formation choices, etc., that are required
(sub-plans for achieving this clean-up are not shown).

The plan for {\lstinline{in_formation}} (lines 26-31) is where
most of the work happens.  First, the leader chooses positions for all
the agents in the system and informs them of their position. Then it
informs all the agents of the formation to be adopted and waits for
the other agents to tell it when they have reached their positions.

Lines 33-36 show the code for assigning positions.  While there is an
agent who has no position, and a position in the formation that has no
agent assigned, the leader will assign that agent to that position and
inform the agent of this fact.  Since
{\lstinline{all_positions_assigned}} is an \emph{achieve goal},
the plan continues to be selected until either there are no more
available agents or no more available positions.  The plan for
{\lstinline{inform_start}} (not shown) works in a similar way.

The description in Listing~\ref{code:leoMAS} is simplified from
the actual code used.  In particular we have omitted those parts that
handle messages from follower agents, which report failures and
aborts; we will not consider these aspects in the discussion of
verification here.

\begin{lstlisting}[float,caption=Multi-Agent LEO System: Leader Agent, style=easslisting, language=Gwendolen,label=code:leoMAS]
:Belief Rules:

B all_positions_assigned(Formation) :- 
          lnot (B pos(Formation, Pos), lnot (B position(Ag, Pos)));
B in_formation(F) :- lnot (B pos(F, P), lnot(B agent_at(P)));
B agent_at(Pos) :-   B position(Ag, Pos), B maintaining(Ag);
B some_formation :- B desired_formation(F1), B in_formation(F1);

B desired_formation(line) :- B one_formation;

B desired_formation(square) :- lnotB one_formation;
B desired_formation(line) :- lnotB one_formation, B in_formation(square);

:Plans:

+! some_formation [achieve] : {lnot B formation(F), B desired_formation(Form)} <-
  +! in_formation(Form) [achieve];
+! some_formation [achieve] : {B formation(F), B desired_formation(F)} <-
  +! in_formation(F) [achieve];
+! some_formation [achieve] : {B formation(F1), lnot B desired_formation(F1), 
                               B desired_formation(Form1)} <-
  +!cleanup_initialisation(F1) [perform],
  +!cleanup_formation(F1) [perform],
  +! in_formation(Form1) [achieve];

+! in_formation(F) [achieve] : {True} <-
  +formation(F),
  +! all_positions_assigned(F) [achieve],
  +! inform_start [achieve],
  *in_formation(F),
  +! cleanup_initialisation(F) [perform] ;

+! all_positions_assigned(Formation) [achieve] : 
  {B agent(Ag), lnotB position(Ag, X), B pos(Formation, Y), lnotB position(Ag2, Y)} <-
  .send(Ag, :tell, position(Y)),
  +position(Ag, Y);

// Information or Requests from other agents
+ aborted(Reason, Ag) : 
   {B position(Ag, X), G some_formation [achieve], ~B maintaining(Ag)}  <-
  +.lock,
  -position(Ag, X),
  -informed(Ag, F),
  -.lock,
  .send(Ag, :perform, new_instruction(drop_formation(F))),
  -! some_formation [achieve];
			
+! send_position(Ag) [perform] : {B position(Ag, X)} <-
   .send(Ag, :tell, position(X));
+! send_position(Ag) [perform] : {~ B position(Ag, X)} <-
   .send(Ag, :tell, position(none));
						
// Plans for cleaning up after a formation is achieved.  
+! formation_clear(F) [achieve] : {B pos(F, P), B position(Ag, P)} <-
  -position(Ag, P);
+! agent_pos_clear [achieve] : {B maintaining(Ag)} <-
  -maintaining(Ag);
+! informed_clear(F) [achieve] : {B informed(Ag, F)} <-
  .send(Ag, :perform, drop_formation(F)),
  -informed(Ag, F);
\end{lstlisting}

\section{Adaptive Cruise Control: The Abstraction Engines}
\label{adaptive:abs}
Listing~\ref{code:abs_single} shows the abstraction engine for our simple example where a car attempts to drive a the speed limit in a single lane.  As sensor input the abstraction engine receives the speed of the car and the acceleration dictated by pressure on the acceleration or brake pedals.  It is also informed whether it is currently a safe distance from the car in front.  These inputs are passed on to the Reasining Engine as shared beliefs in lines 1-14.  Lines 16-20 handle instructions from the rational engine.  Where the driver is using the acceleration or brake pedal the driver's values are used for acceleration or braking, otherwise a simple \lstinline{accelerating} or \lstinline{braking} command is used and a random value invoked in the simulation.

\begin{lstlisting}[float,caption=Cruise Control Agent (Single Lane):Abstraction Engine, style=easslisting, language=Gwendolen,label=code:abs_single]
:Plans:

+safe_in_lane: {lnotB safe} <- assert_shared(safe);
-safe_in_lane: {B safe} <- remove_shared(safe);

+speed(S) : {B speed_limit(Y), lnotB at_speed_limit, Y < S} <- 
   assert_shared(at_speed_limit);
+speed(S) : {B speed_limit(Y), B at_speed_limit, S < Y} <- 
   remove_shared(at_speed_limit);

+acceleration_pedal(A) : {True} <- assert_shared(driver_accelerates);
+brake_pedal(B) : {True} <- assert_shared(driver_brakes);
-acceleration_pedal(A) : {True} <- remove_shared(driver_accelerates);
-brake_pedal(B) : {True} <- remove_shared(driver_brakes);

+! brake [perform]: {B brake_pedal(B)} <- brake(B);
+! accelerate [perform]: {B acceleration_pedal(A)} <- accelerate(A);
+! brake [perform]: {lnotB brake_pedal(B)} <- braking;
+! accelerate [perform]: {lnotB acceleration_pedal(A)} <- accelerating;
+! maintain_speed [perform] : {True} <- accelerate(0);
\end{lstlisting}

Listing~\ref{code:abs_change} shows the abstraction engine used in the lane changing example.  The sensors now provide information about which lane car is in, the distance to the cars in front in both the current lane and the lane to the left, and whether the car is currently crossing lanes (i.e., maneuvering from one lane to another).  The agent has an initial belief about close a car should be to justify initiating an overtaking maneouvre.  

\begin{lstlisting}[float,caption=Cruise Control Agent (Changing Lanes):Abstraction Engine, style=easslisting, language=Gwendolen,label=code:abs_change]
:Initial Beliefs:

// yards
overtaking_at(200)

:Plans:

+lane(0) : {lnotB crossing_lanes} <- assert_shared(in_leftmost_lane);
+lane(I) : {lnotB crossing_lanes, B rightmost_lane(I)} <- assert_shared(in_rightmost_lane);
-lane(0) : {B in_leftmost_lane} <- remove_shared(in_leftmost_lane);
-lane(K) : {B rightmost_lane(K), B in_rightmost_lane} <- remove_shared(in_rightmost_lane);

-crossing_lanes : {B lane(0)} <- assert_shared(in_leftmost_lane), assert_shared(changed_lane);
-crossing_lanes : {B lane(K), B rightmost_lane(K)} <- 
    assert_shared(in_rightmost_lane), 
    assert_shared(changed_lane);
-crossing_lanes : {True} <- assert_shared(changed_lane);

+safe_in_right_lane : {lnotB safe_right} <- assert_shared(safe_right);
+safe_in_left_lane : {lnotB safe_left} <- assert_shared(safe_left);
-safe_in_right_lane : {B safe_right} <- remove_shared(safe_right);
-safe_in_left_lane : {B safe_left} <- remove_shared(safe_left);

+car(D) : {B overtaking_at(K), lnotB car_ahead_in_lane, D < K} <- 
    assert_shared(car_ahead_in_lane);
+car(D) : {B overtaking_at(K), B car_ahead_in_lane, K < D} <- 
    remove_shared(car_ahead_in_lane);
-car(D): {B car_ahead_in_lane} <- remove_shared(car_ahead_in_lane);
+left_car(D) : {B overtaking_at(K), lnotB car_ahead_in_left_lane, D < K} <- 
    assert_shared(car_ahead_in_left_lane);
+left_car(D) : {B overtaking_at(K), B car_ahead_in_left_lane, K < D} <- 
    remove_shared(car_ahead_in_left_lane);
-left_car(D) : {B car_ahead_in_left_lane} <- remove_shared(car_ahead_in_left_lane);

+! change_right [perform]: {B lane(I)} <-
    J = I + 1,
    move_lane(J);
+! change_left [perform]: {B lane(I)} <-
    J = I - 1,
    move_lane(J);
\end{lstlisting}

\section{Logics Workbench Input}
\label{app:lwb}
The Logics Workbench is an interactive system aiming to facilitate the access to logic formalisms.  It can be accessed online from an input shell currently located at \url{http://www.lwb.unibe.ch/shell/}.  

We represented the agent concepts as propositions so $\lbelief{s}{\mathsf{human}}$ became \texttt{bh}, $\lbelief{a}{at(0,0)}$ became \texttt{ba00}, $at(human,0,0)$ became \texttt{ah00}, $at(robot, 0, 0)$ became \texttt{ar00} and so on.   

This allowed us to simply represent the hypotheses and theorems from section~\ref{sec:scen2}.  For instance $\mathsf{found\_human}$ in \eqref{eq:found_human} became

\begin{verbatim}
found_human := bh & ( (ah00 & ar00) v (ah01 & ar01) v (ah02 & ar02) v 
                            (ah10 & ar10) v (ah11 & ar11) v (ah12 & ar12) v 
                            (ah20 & ar20) v (ah21 & ar21) v (ah22 & ar22));
\end{verbatim}

We defined each of the hypotheses in this way and then used \texttt{provable} to prove the theorem.  The sequence of commands used were:
\begin{verbatim}
> load(pltl);
pltl> mc_thm := G ~bh -> (F ba00 & F ba01 & F ba02 & F ba10 & F ba11 & 
                          F ba12 & F ba20 & F ba21 & F ba22);
pltl> found_human := bh & ( (ah00 & ar00) v (ah01 & ar01) v (ah02 & ar02) v 
                            (ah10 & ar10) v (ah11 & ar11) v (ah12 & ar12) v 
                            (ah20 & ar20) v (ah21 & ar21) v (ah22 & ar22));
pltl> correct_sensors := G (bh <-> ( (ah00 & ar00) v (ah01 & ar01) v 
                               (ah02 & ar02) v (ah10 & ar10) v (ah11 & ar11) v 
                               (ah12 & ar12) v (ah20 & ar20) v (ah21 & ar21) v 
                               (ah22 & ar22)));
pltl> correct_movement := G (ba00 <-> ar00) & G (ba01 <-> ar01) & 
                          G (ba02 <-> ar02) & G (ba10 <-> ar10) & 
                          G (ba11 <-> ar11) & G (ba12 <-> ar12) & 
                          G (ba20 <-> ar20) & G (ba21 <-> ar21) & 
                          G (ba22 <-> ar22);
pltl> cond:= G ah00 v G ah01 v G ah02 v G ah10 v G ah11 v 
             G ah12 v G ah20 v G ah21 v G ah22;

pltl> provable( (mc_thm & correct_sensors & correct_movement & cond) -> F found_human);
\end{verbatim}

\noindent We proved~\eqref{eq:combined} in a similar fashion with
\texttt{bsf} representing $\mathbf{B}_s \mathsf{found}$, \texttt{bss}
representing $\lbelief{s}{\mathsf{sent(lifter, human(X, Y))}}$,
\texttt{blr} representing $\lbelief{l}{\mathsf{rec(searcher, human(X,
    Y))}})$, and \texttt{blf} representing $\mathbf{B}_l
\mathsf{free(human)}$.  The sequence of commands used were:

\begin{verbatim}
> load(pltl);
pltl> mc_sthm:= G (bsf  -> F bss);
pltl> mc_lthm:= G (blr -> F blf);
pltl> reliable_comms:= G (bss -> F blr);
pltl> thm10:= G (bsf -> F blf);
pltl> provable( mc_sthm & mc_lthm & reliable_comms -> thm10);
\end{verbatim}

\end{document}